\documentstyle[mncite,psfig]{mn}

\newcommand{\kms}{km\hspace*{0.35ex}s$^{-1}$}
\newcommand{\pcc}{cm$^{-3}$}

\newcommand{\lesssim}{\raisebox{-0.6ex}{$\,\stackrel
{\raisebox{-.2ex}{$\textstyle <$}}{\sim}\,$}}
\newcommand{\gtrsim}{\raisebox{-0.6ex}{$\,\stackrel
{\raisebox{-.2ex}{$\textstyle >$}}{\sim}\,$}}

\title[INTEGRAL spectroscopy of powerful radio galaxies]{INTEGRAL
spectroscopy of three powerful radio galaxies:\\ Jet-cloud
interactions seen in 3-D}

\author[C.\ Sol\'{o}rzano-I\~{n}arrea \& C. N. Tadhunter]
{C. Sol\'{o}rzano-I\~{n}arrea$^{1}$\thanks{E-mail:csi@roe.ac.uk} and 
C. N. Tadhunter$^2$ \\ 
$^{1}$Institute for Astronomy, University of Edinburgh, Royal Observatory, 
Edinburgh, EH9 3HJ, UK\\ 
$^{2}$Department of Physics and Astronomy, University of Sheffield, 
Sheffield, S3 7RH, UK\\ }

\begin{document}
\maketitle

\begin{abstract}

Integral-field spectroscopic observations are presented for three powerful
radio galaxies, namely 3C277.3 (Coma A; z=0.0857), 3C171 (z=0.2384) and
3C265 (z=0.811), which are known to be undergoing jet-cloud interactions.
The morphology, kinematics and ionization of the gas in the emission-line
structure of these sources are mapped and analysed. One-dimensional
spectra are also extracted and integrated over the different emission-line
regions in each galaxy.

In two of the galaxies (3C277.3 and 3C171) the radio sizes are of similar
extent to the emission-line structure. For these, enhanced emission-line
regions are found associated with the radio structures, in addition to
complex kinematics and low ionization states close to the radio hotspots,
indicating that jet-induced shocks disturb and ionize the
gas. Interestingly, the bright --- presumably shock-ionized ---
emission-line region coincident with the radio jet knot in 3C277.3 shows
quiescent kinematics and high ionization state. Possible explanations for
this puzzling result are proposed.

The images of 3C171 and 3C265 indicate that the lateral expansion of the
cocoon has a significant effect on the kinematics and ionization of the
gas, showing for the first time that the effects of the radio source are
felt far from the jet axis.

In addition, the presence of a stellar-photoionized HII region is detected
in the extended emission-line nebula of the radio galaxy 3C277.3.

\end{abstract}


\section{Introduction}

\begin{table*}
\begin{center}
\begin{tabular}{lcccccccc}\hline \\
{\bf Source}&{\bf Redshift}&{\bf P$_{178MHz}$}&{\bf Ang. scale}&{\bf Radio size}&{\bf
 Neb. size} & {\bf Radio PA}&{References}\\
 & & (W Hz$^{-1}$)& (kpc/arcsec) & (kpc) & (kpc) & (deg.)&\\ \\ \hline \\
3C277.3 & 0.0857 & 2.86$\times$10$^{26}$& 2.15 & 96 & 96 & 155 & bri81,vbre85  \\
3C171 & 0.2384 & 5.15$\times$10$^{27}$& 4.76 & 55 & 67 & 100 & cla98\\
3C265 & 0.8110 & 7.10$\times$10$^{28}$& 8.25 & 643 & 289 & 107 &fer93,sol02\\ \\ \hline
\end{tabular}
\caption[]{Properties of the three radio galaxies in the sample. The
redshift is given in column 2. Column 3 shows the radio power at 178 MHz,
calculated from the 178-MHz flux densities and spectral indexes published
in \scite{spinrad85}, for the cosmology adopted in this paper (H$_{0}$ =
50 km s$^{-1}$ Mpc$^{-1}$ and $\Omega_{0}$~=~1). The angular scale for the
same cosmology is given in column 4. The following columns give the extent
of the radio source along the radio axis, the largest nebular size and
the position angle of the radio axis for each source, with the references to
these values listed in the last column.  References: bri81 --
\pcite{bridle81}; cla98 -- \pcite{clark98}; fer93 -- \pcite{fernini93};
sol02 -- Sol{\'o}rzano-I{\~n}arrea et al. 2002; vbre85 --
\pcite{vanbreugel85}.}
\end{center}
\label{proptab}
\end{table*}

Radio galaxies are amongst the most massive galaxies in the early Universe
(e.g. \pcite{best98,vanbreugel98,ivison98,mclure2000}), and as such are
unique cosmological probes of the formation and evolution of these giant
ellipticals. Further, because the active galactic nuclei (AGN) of radio
galaxies are obscured, the host galaxies can be studied directly.

In the currently popular hierarchical growth models, galaxies are built up
through mergers of smaller sub-units \cite{white91}, and such galaxy
mergers have been proposed to trigger the radio source activity
(e.g. \pcite{heckman86,simpson02}). This merger theory is supported by the
fact that many powerful low-redshift radio galaxies present peculiarities
in their morphology, such as tails, shells, etc. \cite{heckman86}.  The
importance of mergers in triggering high-redshift radio galaxies remains
unclear, however.

A remarkable feature of some powerful radio galaxies is that they possess
luminous extended emission-line regions (EELR), which can extend up to
$\sim$150 kpc from their nuclei, and are usually aligned along the radio
axis (e.g. \pcite{mccarthy87}). The ionization of these EELR is thought to
be due to a combination of photoionization by the central AGN and shocks
induced by interactions with the radio source, with AGN-photoionization
dominating near the nucleus, and jet-cloud interactions becoming more
important at larger radii (e.g. \pcite{carmen02}). By investigating the
properties of these EELR we can learn about the interactions between the
radio source and its environment. Moreover, the properties of the
intrinsic (pre-shocked) gas can provide important clues about the origin
of this extended gas, and hence also about the formation and evolution of
these massive galaxies and the triggering of the AGN.

The great majority of the spectroscopic studies carried out to investigate
the kinematics and physical conditions of the emission-line gas in radio
galaxies are based on long-slit observations, which usually concentrate on
the jet-induced shocked structures along the radio axis. These studies
give a very limited view of the EELR properties due to the lack of
knowledge away from the radio axis. In order to learn about the origin of
the extended gas and the triggering of the nuclear activity, as well as
the evolution of the host galaxies, the properties of the intrinsic
gas also need to be studied. Full maps of both kinematics and
ionization of the extended haloes are required in order to separate the
intrinsic gas properties from those resulting from the interactions with
the radio source.

In this paper integral-field spectroscopic observations are presented of a
pilot sample of three powerful radio galaxies at different redshifts, all
of which are known to be undergoing jet-cloud interactions at some
level. These are 3C277.3 (z=0.0857), 3C171 (z=0.2384) and 3C265
(z=0.811). Properties of each source are listed in
Table~\ref{proptab}. With these observations the velocity field, linewidth
and ionization structures of the emission-line gas in these sources are
studied in two spatial dimensions, allowing the AGN phenomenon to be
further understood.

The paper is arranged as follows. In Section 2, a summary is provided of
previous observations of the three sources in the sample. Section 3
contains details of the new observations, data reduction and analysis. The
results are presented in Section 4 and discussed in Section 5. Conclusions
are summarised in Section 6.

Throughout this paper a Hubble constant of H$_{0}$ = 50 km s$^{-1}$
Mpc$^{-1}$ and a density parameter of $\Omega_{0}$~=~1 are assumed.
The resulting angular scales of the galaxies in the sample are given in
Table~\ref{proptab}.

\section{Previous observations}

\begin{table*}
\begin{center}
\begin{tabular}{ccclccccc}\hline \\
{\bf Source}&{\bf t$_{exp}$}&{\bf CCD}&{\bf \ Grat.}&{\bf$\lambda$ Range $^{(*)}$}&{\bf Pix. scale}&
 {\bf Resolution}& {\bf Airm.} &{\bf Seeing} \\
  & (s) & & & (\AA) & (\AA) & (\AA) & & (arcsec)\\ \\ \hline \\
3C277.3 -- Central & 3 $\times$ 1800& Tek6& R600B& 4596 -- 7591 &2.98& 4.5 -- 6.2 & 1.02 & 1.5\\
3C277.3 -- Northern& 3 $\times$ 2000& Tek6& R1200B& 6324 -- 7713&1.40& 2.2 -- 2.7& 1.38 & 1.4\\
3C171 & 6 $\times$ 1800 & Tek6 & R600B & 4596 -- 7591 & 2.98 & 4.5 -- 6.2 & 1.23 & 1.5 \\
3C265 & 3 $\times$ 1800 & Tek6 & R600B & 4596 -- 7591 & 2.98 & 4.5 -- 6.2 & 1.02 & 1.5\\ \\ \hline
\end{tabular}
\caption{Log of the INTEGRAL spectroscopic observations. $^{(*)}$ Common
wavelength range covered by all the fibres; the lower and upper limits of
the wavelength range covered by each single fibre can vary up to --50
and +50 \AA \ of those shown in the table, respectively.  }
\end{center}
\label{logtab}
\end{table*}

\subsection{3C277.3 (Coma A)}

\begin{figure}
\centerline{ \psfig{figure=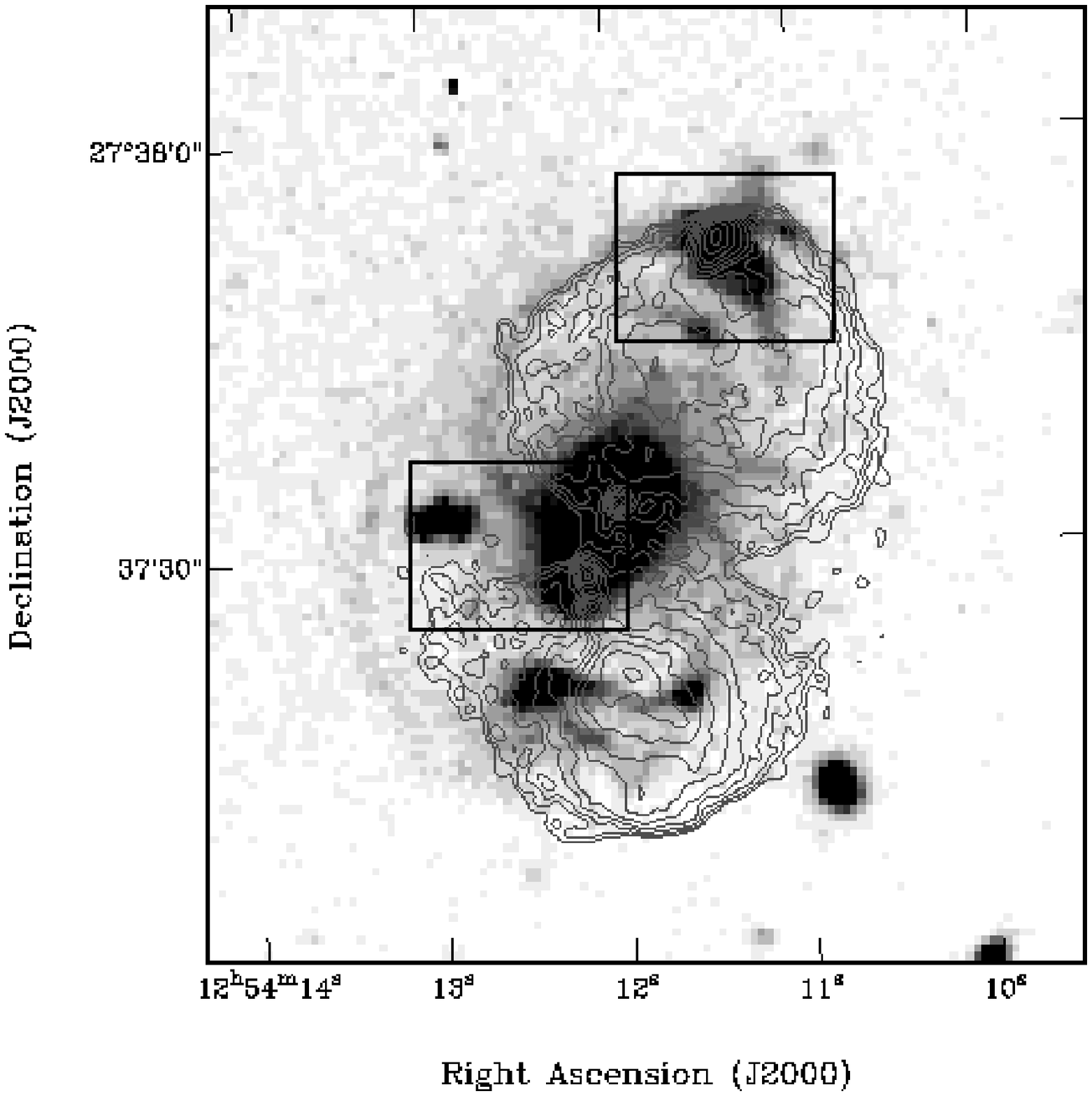,angle=-90,width=8cm}}
\caption[]{H$\alpha$ image of Coma A (from \pcite{tadhunter2000}) overlaid
with contours of radio emission at 21 cm (from \pcite{vanbreugel85}). The
two boxes indicate the regions we have observed with INTEGRAL.}
\label{comatot+rad}
\end{figure}

The radio source 3C277.3, at a redshift z=0.0857, has an intermediate
morphology between FRI and FRII \cite{fanaroff74} structures (see radio
map in \pcite{vanbreugel85}, reproduced in Fig.~\ref{comatot+rad}). The
radio emission, mainly consisting of two wide, diffuse lobes, has an
extent of $\sim$ 45 arcsec (96 kpc) along the radio axis. Whilst the
northern radio lobe contains a resolved bright hot spot at its outer
northern edge, the southern lobe presents two bright jet knots extending
southward from the radio core.

Detailed radio and optical studies of 3C277.3 by \scite{vanbreugel85} show
evidence for an interaction between the radio source and the ambient
gas. There is a strong morphological association between the emission-line
gas and radio structures, and the emission-line kinematics appear to be
influenced by the radio source. Moreover, the deflection and lighting up
of the southern radio jet near a region of bright line emission suggests
that the radio jet has collided with a massive cloud which decollimates
the jet and is partially entrained in it \cite{vanbreugel85}. This is
clearly seen in Hubble Space Telescope (HST) images of this galaxy
\cite{martel99,capetti2000} which show a filamentary structure at the
location of the bright knots.

Deep narrow-band H$\alpha$ images of 3C277.3 (see Fig.~\ref{comatot+rad})
reveal a spectacular system of interlocking emission-line arcs and
filaments, which extend almost as far perpendicular as parallel to the
radio axis \cite{tadhunter2000}. These authors suggest that the close
match between the radio and emission-line morphologies in 3C277.3 may
indicate that not only do direct interactions with the radio jet ionize the
gas, but also the expansion of the radio lobes into the halo plays an
important role. Recently, \scite{morganti02} have detected HI absorption
against both radio lobes of 3C277.3, at large distances from the nucleus
($\sim$ 30 kpc), suggesting that the radio lobes are expanding into a
large gas disk. This is supported by the morphology of the ionized gas and
by the good match between the velocities of the neutral hydrogen and those
of the extended ionized gas, which indicates that they are two phases of
the same disk-like structure of at least 60 kpc in diameter.

\subsection{3C171}

\begin{figure}
\centerline{\psfig{figure=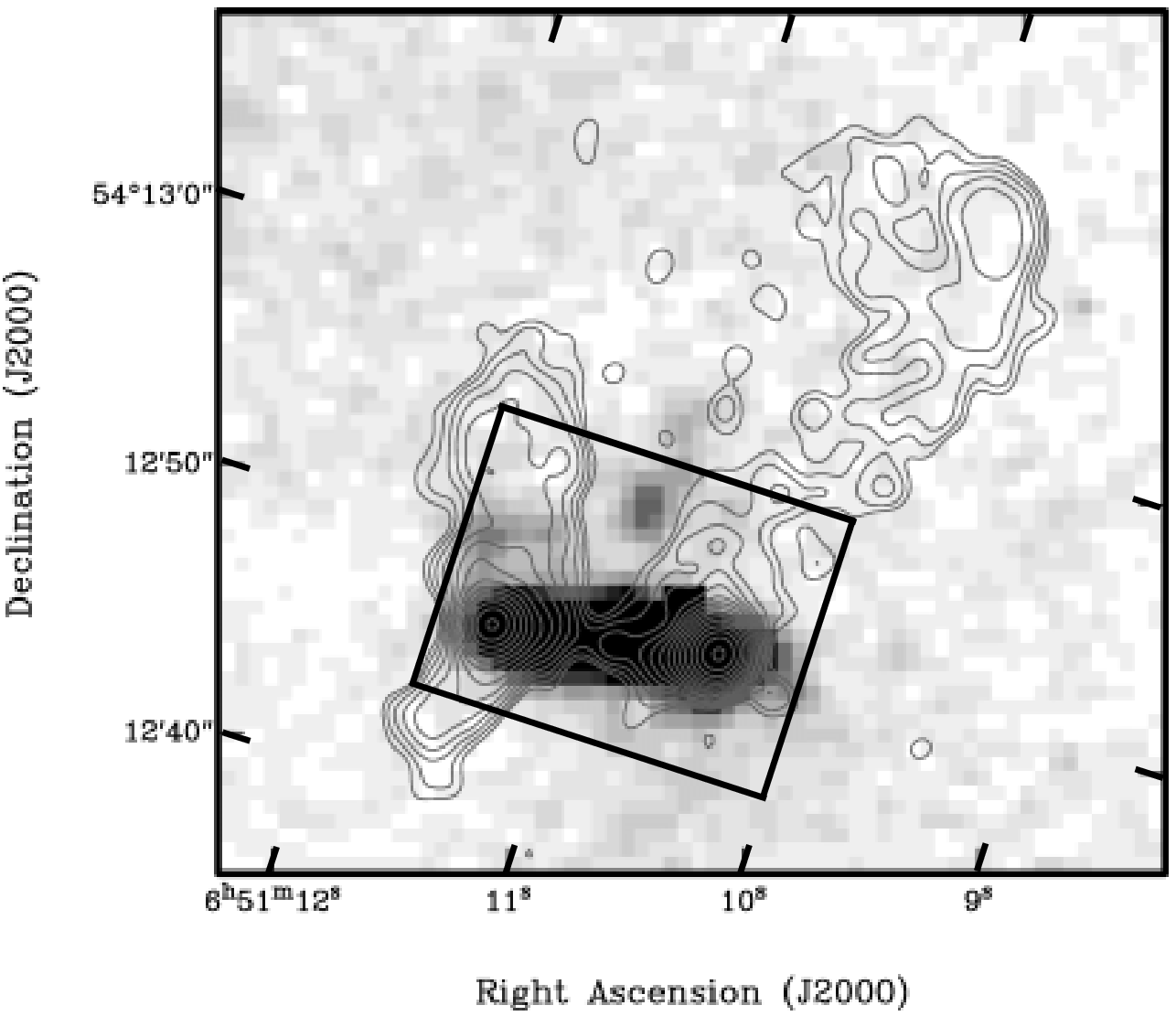,width=8cm}}
\caption[]{H$\alpha$ image of 3C171 (from \pcite{tadhunter2000}) overlaid
with contours of radio emission at 6 cm (from \pcite{blundell96}). The box
indicates the region we have observed with INTEGRAL.}
\label{171rad}
\end{figure}

3C171 is a powerful radio source at a redshift z=0.2384. Its radio
emission has an unusual morphology. While at high surface brightness
levels the source has a normal FRII \cite{fanaroff74} structure, with an
extent of $\sim$ 11.5 arcsec (55 kpc), at lower surface brightness levels
(see Fig.~\ref{171rad}) long plumes are seen associated with each hotspot
and extending up to 20 arcsec north and south perpendicular to the radio
axis (see radio maps in \pcite{blundell96} and \pcite{hardcastle97}).

Optical images show line emission extending along the radio axis of 3C171,
with an spatial extent similar to that of the high-surface-brightness
radio emission structure \cite{heckman84,baum88}. The first detailed study
of 3C171 (Heckman et al. 1984)\nocite{heckman84} provided evidence for an
interaction between the radio source and the ambient gas of the host
galaxy: complex emission-line kinematics along the radio axis were
revealed by the long-slit spectra, showing the highly perturbed state of
the gas.  These results are supported by a more recent detailed study of
this source, based on high-resolution long-slit spectroscopic observations
\cite{clark98}. These data reveal close radio-optical associations, strong
line splittings and underlying broad kinematics components in the extended
gas along the radio axis, a general low ionization state of the
line-emitting gas, ionization minima coincident with the radio hotspots,
and an anticorrelation between linewidth and ionization state in the
extended gas. These all provide strong evidence that the morphology,
kinematics and physical conditions of the emission-line gas in 3C171 are
defined by jet-induced shocks. In addition, integral-field studies of this
galaxy over a short wavelength range containing H$\beta$ and
[OIII]$\lambda$5007 show emission-line ratios and kinematics which are
consistent with a jet-cloud interaction scenario in the extended gas of
3C171 near the radio hotspots \cite{marquez2000}.

Whilst the brightest emission-line structures in 3C171 are closely aligned
with the radio axis, the existence of low-surface-brightness features in
the direction perpendicular to the radio axis is revealed by a deep
narrow-band imaging study of 3C171 (Fig.~\ref{171rad};
\pcite{tadhunter2000}). These authors argue that these fainter emission-line
regions lying away from the radio axis represent intrinsic structures of
the gas, whose association with the extended radio structures provide
evidence that they are ionized by their interaction with the radio lobes.

\subsection{3C265}

3C265 is a large (78 arcsec; 643 kpc) FRII \cite{fanaroff74} radio source,
at a redshift z=0.811 (see radio map in \pcite{fernini93}). It has an
extreme emission-line luminosity \cite{thesismccarthy88}, which makes
3C265 ideal for the study of emission-line properties in distant radio
galaxies.  Its continuum structure extends over more than 10 arcsec and
[OII]$\lambda$3727 images show emission over more than 30 arcsec
(\pcite{rigler92,mccarthy95}; Sol{\'o}rzano-I{\~n}arrea et
al. 2002)\nocite{carmen02}. HST images reveal the bizarre optical
morphology of this object, with the nucleus surrounded (in projection) by
emission regions or companion galaxies \cite{longair95}.

\begin{figure}
\vspace*{0.3cm}
\centerline{\psfig{figure=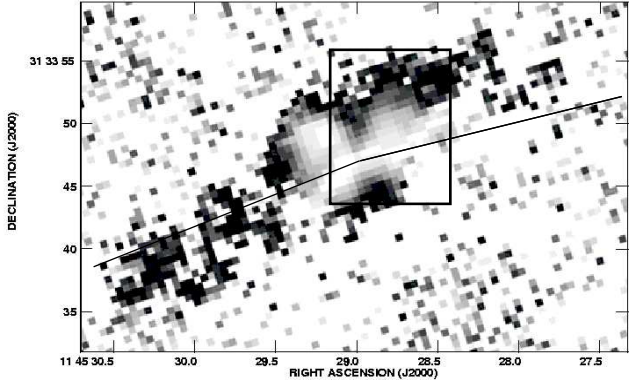,angle=-90,width=8.5cm}}
\caption[]{Image of 3C265 showing a map of the [OII]/[OIII] line ratio (from
Sol{\'o}rzano-I{\~n}arrea et al. 2002). White indicates high ionization
and black indicates low ionization. Near the nucleus the ionization
structure has a `butterfly' shape, consistent with the bicone predicted by
the unified schemes \cite{barthel89}. The box indicates the region we have
observed with INTEGRAL.} 
\label{cones265}
\end{figure}

Spectroscopic and polarimetric studies of 3C265
\cite{jannuzi91,dey96,di-serego-ali96} clearly indicate that the extended
gas is being illuminated by an obscured powerful quasar in the centre of
the galaxy. In addition, the near-UV emission-line ratios of the extended
gas in 3C265 are consistent with AGN-photoionization
\cite{best2000}. These results are supported by recent deep narrow-band
images of 3C265 (Sol{\'o}rzano-I{\~n}arrea et al. 2002)\nocite{carmen02},
shown in Fig.~\ref{cones265}, which reveal the existence of a
$\sim$60$^{\circ}$ half-opening angle illumination bicone near the nucleus
(r$\lesssim$60 kpc). This is predicted by the unified schemes for radio
sources \cite{barthel89}, whereby the AGN in radio galaxies is obscured by
a surrounding dusty torus which defines the opening angle of the
ionization cones. This result indicates that anisotropic illumination from
the central AGN dominates on a small scale in 3C265.  By contrast, the
close alignment with the radio axis of low-ionization material at larger
distances from the nucleus (r$\gtrsim$80 kpc), suggests that jet-cloud
interactions may become the dominant mechanism of the emission-line gas on
larger scales. Further, the presence of high-velocity gas close to the
radio axis, at $\sim$20 kpc from the nucleus (\pcite{tadhunter91};
Sol{\'o}rzano-I{\~n}arrea et al. 2002), shows that the kinematic effects
of jet-induced shocks are still important in the near-nuclear
regions.\nocite{carmen02}

Although on a larger scale the emission-line structures in 3C265 appear to
be closely aligned along the radio axis, near the nucleus the
highest-surface brightness structure is elongated and aligned with the
closest companion galaxy, and misaligned by approximately 35$^{\circ}$
relative to the radio axis, which is still consistent with the large
half-opening angle of the ionization bicone. This indicates that much of
the emission-line structure  reflects the intrinsic distribution of the
gas, and suggests that the material present along the direction of the
optical/UV elongation is associated with an interaction/merger involving
3C265 and this companion (Sol{\'o}rzano-I{\~n}arrea et
al. 2002).\nocite{carmen02}

\section{Observations, reduction and analysis}

Integral-field spectroscopic observations were carried out on the night of
2001 January 16 for the northern region of 3C277.3, in service mode, and
on the night of 2001 January 21 for 3C265, 3C171 and the `central' region
of 3C277.3, using the INTEGRAL spectrograph on the 4.2-m William Herschel
Telescope (WHT) on La Palma (Spain). The fibre bundle SB2 was used, which
has a fibre core diameter of 0.9 arcsec, and provides an effective field
of view of 14.6 $\times$ 11.3 arcsec$^{2}$. Only in the case of 3C171
could the entire emission-line nebula be covered in a single field
exposure. The regions observed for each source are indicated with boxes in
Figs.~\ref{comatot+rad}, \ref{171rad} and \ref{cones265}. The central
region observed in 3C277.3 is offset from the core so as to cover in the
same exposure the eastern filament $\sim$12 arcsec east of the nucleus and
the emission-line knot $\sim$6 arcsec to the south of the nucleus, in
addition to the nucleus itself.  Details of the observations are presented
in Table~\ref{logtab}.

\subsection{Data reduction}

The reduction of the multi-spectra was carried out using standard tasks
and specific \emph{integral} tasks within the {\small IRAF} software
package and was performed in several stages, as explained in the {\small
INTEGRAL} data reduction manual.  After the bias subtraction, the
apertures corresponding to each fibre were defined and traced, using a
flat-field image.  Following this, all frames were corrected for scattered
light from the spectrograph and flat-fielded. The apertures in each frame
were then extracted into one-dimensional spectra, one per optical
fibre. Using sky flat-fields, the spectra were corrected for the fibre
throughput. The different exposures of each galaxy were then combined,
improving the signal-to-noise and removing the cosmic rays. After that,
the wavelength calibration was performed using both sky lines and
comparison lamps (CuAr and CuNe). The data were corrected for atmospheric
extinction and then flux calibrated (see below) using observations of the
spectrophotometric standard stars SP0804+751 and SP1036+433, which were
obtained on the service night and second run night, respectively.
Following the flux calibration, the galaxy data were sky-subtracted by
using the sky fibres and emission-free regions of the frames.

\subsubsection{Flux calibration}

\begin{table}
\begin{center}
\begin{tabular}{cll}\hline \\
 {\bf Source}&{\bf Line}&{\bf \ Surf. Bright. cutoff} \\
 & &(erg cm$^{-2}$ s$^{-1}$ arcsec$^{-2}$)\\ \\ \hline \\
3C277.3 -- Central & $[$OIII$]$5007&\hspace*{0.7cm}$\sim 8.0 \times 10^{-17}$ \\
\vspace*{0.3cm}
\hspace*{-0.1cm}3C277.3 -- Central & H$\alpha$ &\hspace*{0.7cm}$\sim 5.4 \times 10^{-17}$ \\
\vspace*{0.3cm}
 3C277.3 -- Northern& H$\alpha$ &\hspace*{0.7cm}$\sim 5.5 \times 10^{-17}$ \\ 
 3C171 & $[$OII$]$3727&\hspace*{0.7cm}$\sim 12.5 \times 10^{-17}$ \\
 3C171 & H$\beta$ &\hspace*{0.7cm}$\sim 9.2 \times 10^{-17}$ \\ 
\vspace*{0.3cm}
\hspace*{-0.1cm}3C171 &$[$OIII$]$5007&\hspace*{0.7cm}$\sim 9.2 \times 10^{-17}$ \\ 
 3C265 &$[$OII$]$3727&\hspace*{0.7cm}$\sim 9.5 \times 10^{-17}$ \\ \\ \hline
\end{tabular}
\caption{Surface brightness cutoff (5$\sigma$ level) applied to each
galaxy for the emission lines listed.}
\end{center}
\label{s5tab}
\end{table}

The images of the standard stars were taken greatly out of focus in order
to spread the flux of the star over several fibres, and thus allow to
better estimate the flux loss between fibres. The sky-subtracted
multi-spectra of each standard star were first added up to form a 1-D
spectrum for each star. These were used to correct for the instrumental
sensitivity at each wavelength and to obtain a rough absolute flux
calibration. Given that some flux is lost between the fibres, an accurate
absolute flux calibration is difficult to obtain. This effect was accounted for
in the following way. Firstly, images of each (flux calibrated) star in
several wavelength ranges were reconstructed (using the task
\emph{imarec}). The measured total flux in those wavelength ranges was
then compared to the tabulated flux of each star within those ranges.  The
fraction of overestimated flux, due to the loss of flux between fibres,
was determined for each star, and the corresponding galaxy frames of
either night were then corrected.  Since only one standard star was
observed per night, errors in the absolute flux calibration could not be
derived directly from our INTEGRAL data. To estimate the uncertainty in
the flux calibration, we compared our data with the existing long-slit
spectroscopic observations of the same objects. It was found that the
absolute fluxes obtained from the INTEGRAL data agreed with those derived
from the long-slit spectra to within $\pm$20 \%. Likewise, by comparing
the emission-line ratios derived from our data with those derived from
existing long-slit spectra, we estimate that the relative fluxes are
accurate to well within 10\%.

\subsection{Data analysis}

\subsubsection{Emission-line intensity and line-ratio maps}

Intensity maps for the different emission lines were reconstructed from
the reduced multi-spectra frames, by using the task \emph{imarec}.  The
resulting images were continuum-subtracted in the following way: two
images of the continuum emission in two line-free regions adjacent to the
emission line on either side, were reconstructed and added together. The
resulting continuum image was scaled to the emission-line image, so both
images had the same wavelength bandpass. The scaled continuum image was
then subtracted from the emission-line image.

All pixels with values less than approximately 5$\sigma$, where $\sigma$
is the root-mean-squared variation in the source-free regions of the
image, were blanked in the continuum-subtracted emission-line images. The
surface brightness corresponding to the cutoff values are listed in
Table~3. Line-ratio maps were then obtained by dividing the relevant
emission-line intensity maps by each other. As a checking procedure, the
line ratios were also derived from Gaussian fits of spectra in individual
fibres.

The data have not been corrected for Galactic reddening given that none of
the objects in our sample have low Galactic latitude, and thus the
extinction due to dust in our Galaxy is not large [0.012 $\leq$ E(B-V)
$\leq$ 0.054 mag; taken from NASA/IPAC Extragalactic Database (NED)].

\subsubsection{Velocity field and linewidth maps}

Velocity shifts and linewidths (FWHM) were obtained by both Fourier
cross-correlation and Gaussian fitting\footnote{Although the emission-line
profiles are likely to be complex in some objects (e.g. 3C171), the lines
were fitted with single Gaussians because the S/N was not sufficiently
high in all fibres for a multiple Gaussian fit. } of the emission-line
profiles (1-D spectra, one per fibre, were extracted from the
reduced multi-spectra frames of the galaxies, and then analysed using the
IRAF and DIPSO packages). Results from both methods were consistent with
each other.  The velocity shifts are referred to the velocity of the
line-emitting gas at the continuum centroid of each galaxy. The measured
linewidths were corrected for the spectral resolution of the instrument,
which was derived by using the night-sky emission lines. The instrumental
widths were measured for each fibre individually, since the spectral
resolution changes over the CCD due to the variation of the spectrograph
focus over the detector. The ranges of variation of the instrumental
resolution are listed in Table~\ref{logtab}.  Velocity fields and
linewidths were then mapped for each galaxy. The spatial distributions of
these parameters were reconstructed, by using the task \emph{int-map},
from ASCII files containing the values of the parameters for each
fibre. Those pixels blanked in the emission-line intensity maps, because
they had a S/N $\lesssim$ 5 (see above), were also blanked in the velocity
and linewidth maps.

\subsubsection{Comparison with previous long-slit spectroscopic studies}

\begin{figure}
\centerline{\psfig{figure=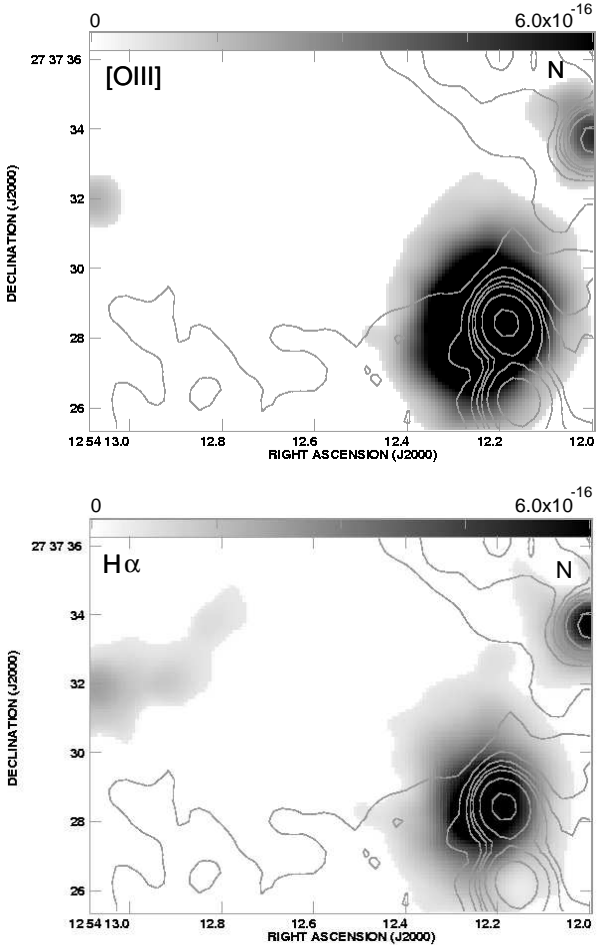,width=8.1cm}}
\caption[]{INTEGRAL grey-scale images of the central region of 3C277.3 in
the [OIII]5007 (top) and H$\alpha$ (bottom) lines, with the continuum
emission subtracted. The `N' indicates the nucleus. The contours of
the 1.4 GHz radio map of \scite{vanbreugel85} are superimposed. The
grey-scale  values range from 0 to 6.0$\times$10$^{-16}$ erg cm$^{-2}$ s$^{-1}$
arcsec$^{-2}$. Radio contour values are 0.7 mJy beam$^{-1}$ $\times$ (-1,
1, 2, 3, 4, 5, 8, 16, 32, 64). }
\label{comanucfl}
\end{figure}

\begin{figure}
\centerline{\psfig{figure=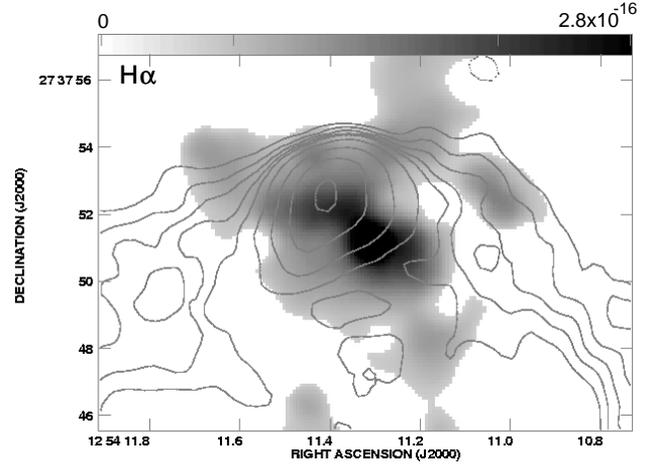,width=8.5cm}}
\caption[]{INTEGRAL grey-scale image of the northern region of 3C277.3 in
H$\alpha$, with the continuum emission subtracted. The contours of
the 1.4 GHz radio map of \scite{vanbreugel85} are superimposed. The
grey-scale  values range from 0 to 2.8$\times$10$^{-16}$ erg cm$^{-2}$ s$^{-1}$
arcsec$^{-2}$. Radio contour values are 0.7 mJy beam$^{-1}$ $\times$ (-1,
1, 2, 3, 4, 5, 8, 16, 32, 64).}
\label{comaNh}
\end{figure}

As a test, the results obtained from the INTEGRAL data were compared with
those obtained from existing long-slit spectra of the three galaxies in the
sample, in the regions of the images with the same PA as the long-slit
spectra. Taking into account the different resolutions and S/N (e.g. the
current data fail to detect the high-velocity cloud in 3C265 probably due
to a lower S/N), the results in the overlapping region are completely
consistent with previous studies. Moreover, these INTEGRAL studies provide a
fuller view of the sources, and also in some cases allow the results to be
interpreted in a way that is impossible with long-slit studies.

\section{Results}

\subsection{3C277.3 (Coma A)}
\label{res:coma}

\subsubsection{Emission-line structure}
\label{res:coma_struc}

Fig.~\ref{comanucfl} presents the continuum-subtracted [OIII]5007  and
H$\alpha$  INTEGRAL images for the `central' region of 3C277.3, with the
contours of the 1.4 GHz radio map of \scite{vanbreugel85}
superimposed. The radio and optical maps were aligned by assuming that the
continuum centroid of the optical emission coincides with the radio core.

From Fig.~\ref{comanucfl} it can be seen that, apart from the bright
region in the nucleus (N), a bright, more extended emission-line region is
located at $\sim$6 arcsec (13 kpc) to the south of the nucleus and is
coincident with the radio jet knot. The close spatial association between
this bright region and the radio knot, together  with the morphology of
the radio source in the southern half (see Fig.\ref{comatot+rad}),
suggests that the southern radio jet has been deflected by a massive
ambient cloud ($\sim$2$\times$10$^{6}$~M$_{\odot}$; \pcite{vanbreugel85}).
This emission-line region is brighter in [OIII] with respect to
the nucleus than in H$\alpha$, indicating that its ionization state is
higher than that of the nucleus.

In addition to the bright regions, a fainter emission-line
structure is detected at $\sim$ 12 arcsec (26 kpc) to the east of the
nucleus. This fainter region, however, does not appear to be associated
with any radio structure; it is located near the end of a
low-surface brightness structure of radio emission, and it probably
represents emission from the intrinsic ambient gas, which has not yet
interacted with the radio source. This region appears to be better
detected in H$\alpha$ than in [OIII], suggesting that it has a low
ionization state. 

Unfortunately, the images of the central region of 3C277.3
(Fig.~\ref{comanucfl}) are not deep enough to have detected the
emission-line arc structure reported by \scite{tadhunter2000}, which
partially lies in the region we have observed with INTEGRAL (see
Fig.~\ref{comatot+rad}).
   
Fig.~\ref{comaNh} presents the continuum-subtracted
H$\alpha$  INTEGRAL image for the northern region of 3C277.3, with the
contours of the 1.4 GHz radio map of \scite{vanbreugel85}
superimposed. The figure shows a bright H$\alpha$ structure in the north
of the galaxy ($\sim$ 20 arcsec north from the nucleus), which is much more
extended than those observed in the central region. This region is
coincident with the northern radio hotspot and seems to be partially
located along the boundaries of the radio source. This close association
between the bright emission-line and radio structures suggests that the
radio source is interacting with the environment at this location.

\subsubsection{Emission-line kinematics}
\label{res:comakinemat}

\begin{figure*}
\centerline{\psfig{figure=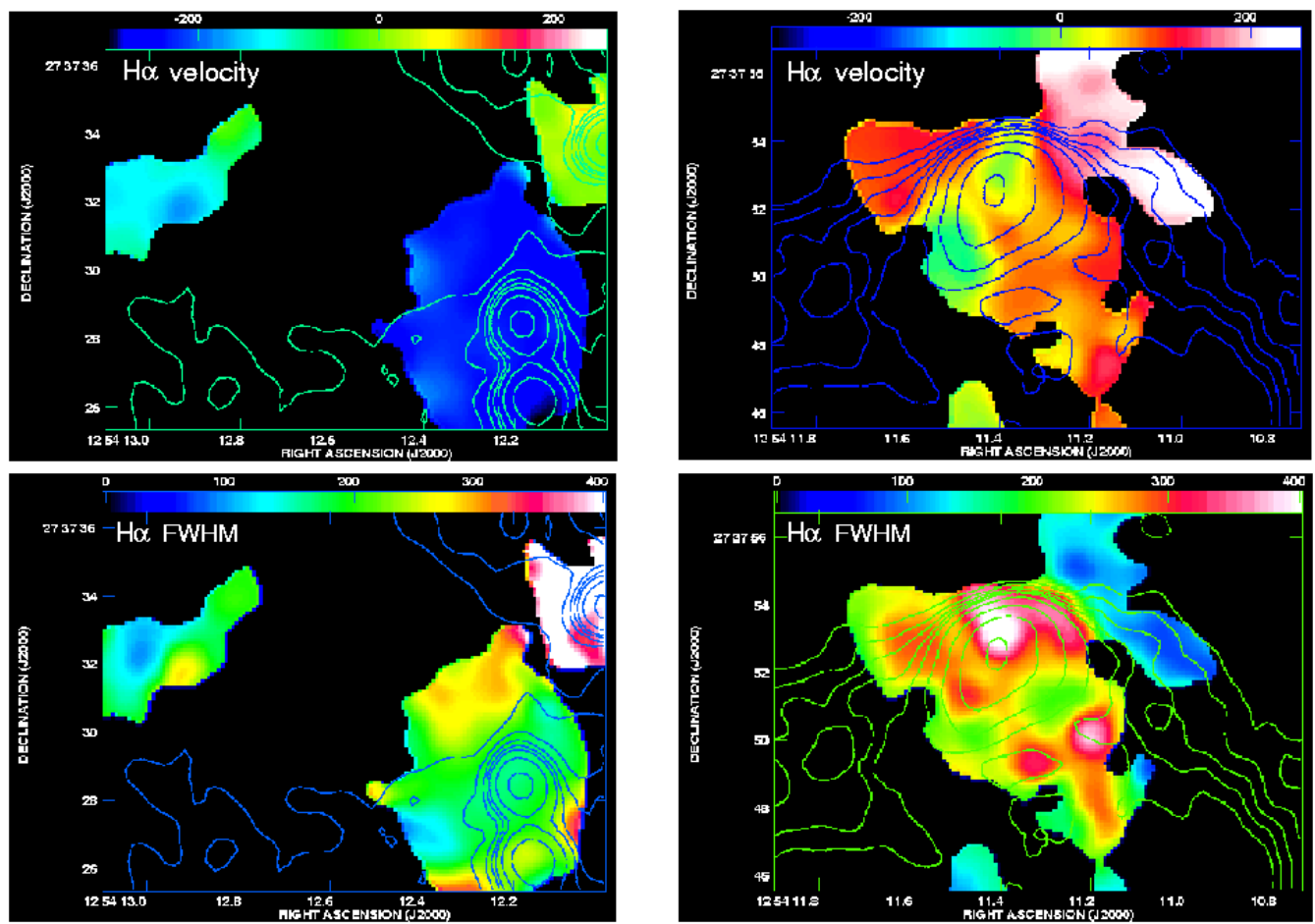,angle=-90,width=17.7cm}}
\vspace{-0.1cm}
\caption[]{H$\alpha$ velocity (top) and instrumentally-corrected
linewidth (bottom) colour maps for the central (left) and northern (right)
regions of 3C277.3. The contours of the 1.4 GHz radio map of
\scite{vanbreugel85} are superimposed. Radio contour levels are as in
Figs.~\ref{comanucfl} and \ref{comaNh}.}
\label{comavelwidth}
\end{figure*}

\begin{figure}
\centerline{\psfig{figure=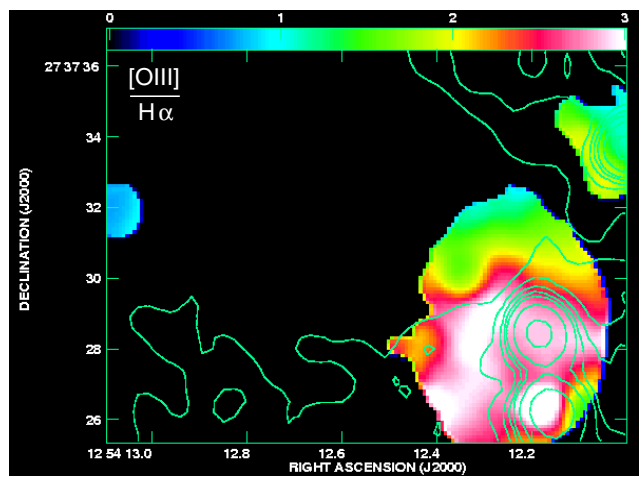,angle=-90,width=8.4cm}}
\vspace*{-0.1cm}
\caption[]{Colour map of the central region of 3C277.3 showing the
[OIII]/H$\alpha$ line ratio. The contours of the 1.4 GHz radio map of
\scite{vanbreugel85} are superimposed. Radio contour levels are as in
Figs.~\ref{comanucfl} and \ref{comaNh}.}
\label{comao3ha}
\end{figure}

Fig.~\ref{comavelwidth} presents the H$\alpha$ velocity field and
(instrumentally-corrected) linewidth colour maps for the two regions
observed in 3C277.3.  Contours of the radio emission at 1.4 GHz (from
\pcite{vanbreugel85}) are superimposed on the colour maps.

The velocity shifts of the emission-line structures in the central region
(Fig.~\ref{comavelwidth}, top left) have an overall amplitude of $\Delta v
\sim$ 200 \kms.  Of particular interest is the fact that the bright region
to the south of the nucleus, and coincident with the radio knot, is
uniformly moving as a whole at a projected velocity of $\sim$ --200 \kms
\ relative to the nucleus.  On the other hand, the velocity of the fainter
structure to the east of the nucleus varies smoothly with an average
velocity of $\sim$ --100 \kms \ relative to the nucleus.

The northern region (Fig.~\ref{comavelwidth}, top right) is, on average,
redshifted with respect to the nucleus, with velocities increasing up to
$\sim$ +250 \kms \ to the north of the radio lobe. Two small regions,
however, located just north and south of the hotspot, appear to be
slightly blueshifted, with velocities of $\sim$ --10 and --50 \kms,
respectively.

\begin{figure}
\centerline{\psfig{figure=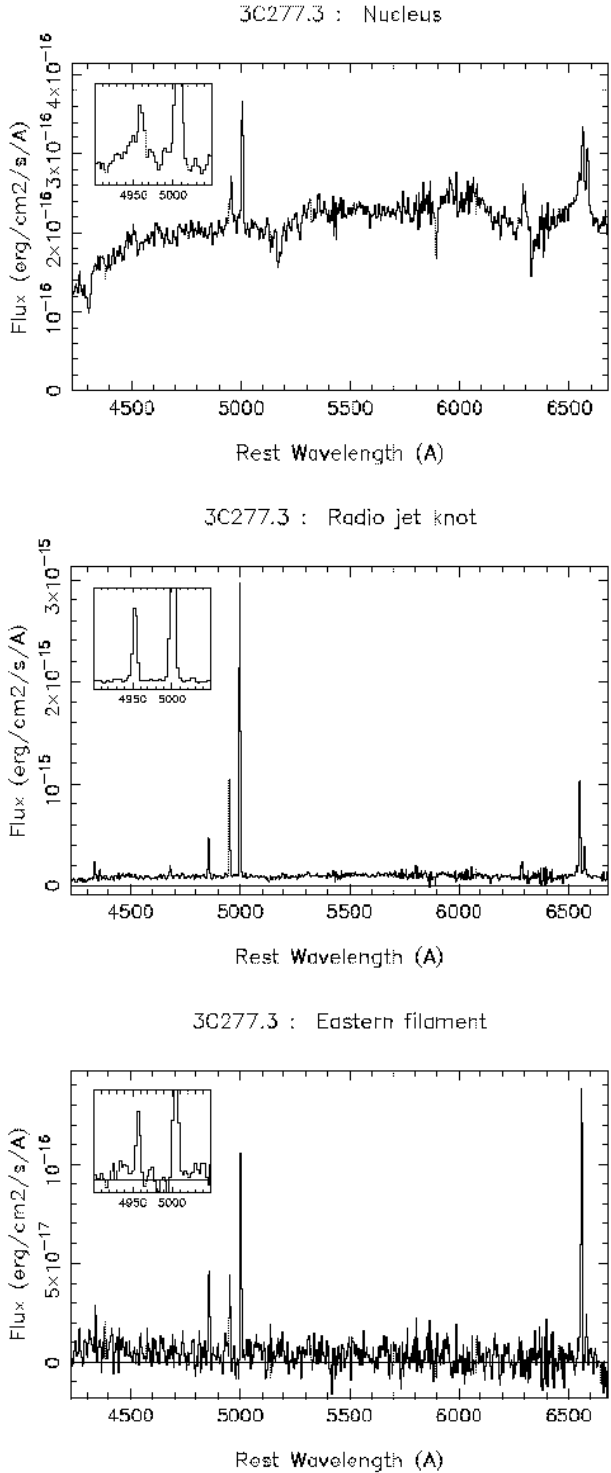,width=8.5cm}}
\caption[]{Integrated spectra for the different regions in 3C277.3:
Nucleus (top: 6.1 arcsec$^2$ aperture), radio knot (middle: 28.2
arcsec$^2$ aperture centred at 6 arcsec south-south-east of the continuum
centroid) and eastern filament (bottom : 7.6 arcsec$^2$ aperture centred
at 12.5 arcsec east of the continuum centroid). The insets show the
spectral region of the [OIII]$\lambda\lambda$4959,5007 doublet.}
\label{speccoma}
\end{figure}

\begin{figure}
\centerline{\psfig{figure=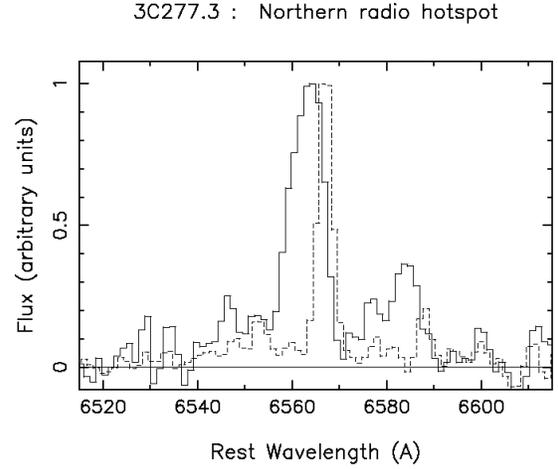,width=7.5cm}}
\caption[]{H$\alpha$ and [NII]$\lambda\lambda$6548,6584 emission-line
profiles for the northern region of 3C277.3. The solid line corresponds to
the region with the broadest lines, 4.6 arcsec$^2$ aperture \ centred at 1
arcsec north-north-west of the hotspot. The dashed line corresponds to the
northernmost region, 10.7 arcsec$^2$ aperture centred at 3.5 arcsec west of
the hotspot. Both spectra are normalised to the peak of the H$\alpha$
line. }
\label{specNcoma}
\end{figure}

The H$\alpha$ linewidth map of the central region
(Fig.~\ref{comavelwidth}, bottom left) shows that the nucleus presents the
broadest emission lines with FWHM $\sim$ 400 \kms. The fainter structure
to the east of the nucleus has an average FWHM of $\sim$ 200 \kms. Of
particular interest is that the bright emission-line region to the south
of the nucleus, which is thought to be undergoing a jet-cloud interaction
\cite{vanbreugel85}, presents narrow lines with FWHM $\sim$ 200 -- 300
\kms; the larger values in this range being found in regions away from the
radio jet knot.  

On the other hand, the H$\alpha$ linewidth map of the northern region
(Fig.~\ref{comavelwidth}, bottom right) shows that the broadest emission
line region (FWHM $\sim$ 430 \kms), with indications of line splitting in
some fibres, is coincident with the radio hotspot. This may indicate that
the gas at this location is perturbed by the radio source.  Moreover, it
is striking how the linewidth suddenly decreases just beyond the hotspot,
reaching values of FWHM $\lesssim$ 100 \kms, suggesting that this is
undisturbed gas that has not yet interacted with the radio-emitting
structures.

\subsubsection{Emission-line ionization state}
\label{res:ionizcoma}

\begin{table*}
\footnotesize
\begin{center}
\begin{tabular}{lcccccccccccc}\hline \\
\ {\large 3C277.3}& \ &\multicolumn{2}{c}{\bf Nucleus}& \ &\multicolumn{2}{c}{\bf Radio
 jet knot}& \ &\multicolumn{2}{c}{\bf Eastern Filament}& \ &\multicolumn{2}{c}{\bf Northern Hotspot}\\
 & & Flux abs. & Flux ratio & & Flux abs. & Flux ratio & &Flux abs. & Flux ratio
 & & Flux abs. & Flux ratio  \\ 
{\bf Line} & & $/ 10^{-15}$ & to H$\beta$& & $/ 10^{-15}$ & to H$\beta$& & $/
 10^{-15}$ & to H$\beta$& & $/ 10^{-15}$ & to H$\beta$ \\ \\ \hline \\
H$\gamma$ & & 
 --- & --- & & 1.39$\pm$0.15 & 0.50$\pm$0.06 & & --- & --- & & --- & --- \\
$[$OIII$]\lambda$4363  & &
 --- & --- & & 0.67$\pm$0.13 & 0.24$\pm$0.05 & & --- & --- & &--- & --- \\
HeII\hspace{0.4ex}$\lambda$4686  & &
 --- & --- & & 0.73$\pm$0.10 & 0.27$\pm$0.04 & & --- & --- & & --- & --- \\
H$\beta$  & &
 --- & $\equiv$1.00 $^{(}$*$^{)}$& & 2.76$\pm$0.10 & 1.00$\pm$0.04 & & 0.35$\pm$0.03 & 1.00$\pm$0.10 & & --- & $\equiv$1.00 $^{(}$*$^{)}$ \\ 
$[$OIII$]\lambda$4959 & &
 0.73$\pm$0.07 & 1.86$\pm$0.19 & & 7.47$\pm$0.08 & 2.70$\pm$0.10 & & 0.26$\pm$0.03 & 0.77$\pm$0.11 & & --- & --- \\
$[$OIII$]\lambda$5007 & &
 1.72$\pm$0.07 & 4.41$\pm$0.27& & 21.25$\pm$0.10 & 7.69$\pm$0.28 & & 0.81$\pm$0.03 & 2.32$\pm$0.24 & & --- & --- \\
$[$OIII$]\lambda$6300  & &
 0.83$\pm$0.15 & 2.09$\pm$0.41 & & --- & --- & & --- & --- & & 0.37$\pm$0.12 & 0.25$\pm$0.10  \\
$[$NII$]\lambda$6548 & &
 0.40$\pm$0.03 & 1.03$\pm$0.12 & & 1.02$\pm$0.15 & 0.37$\pm$0.06 & & 0.10$\pm$0.05 & 0.28$\pm$0.12 & &0.62$\pm$0.15 & 0.40$\pm$0.14 \\
H$\alpha$  & &
 1.20$\pm$0.05 & $\equiv$3.10 $^{(}$*$^{)}$ & & 8.30$\pm$0.15 & 3.00$\pm$0.12 & & 1.11$\pm$0.05 & 3.14$\pm$0.33 & & 4.72$\pm$0.12 & $\equiv$3.10 $^{(}$*$^{)}$ \\
$[$NII$]\lambda$6583  & &
 1.02$\pm$0.05 & 2.61$\pm$0.17 & & 2.52$\pm$0.18 & 0.91$\pm$0.07 & & 0.17$\pm$0.05 & 0.47$\pm$0.13 & & 1.08$\pm$0.13 & 0.71$\pm$0.20 \\ \\ \hline
\end{tabular}
\end{center}
\vspace{-0.2cm}
\caption[]{Emission-line integrated fluxes for the different regions in
3C277.3. Apertures for the nucleus, radio jet knot and eastern filament
are like in Fig.~\ref{speccoma}. The aperture of the northern hotspot
region is 33.6 arcsec$^2$ centred at 1 arcsec south of the hotspot. The
absolute fluxes are in units of erg cm$^{-2}$ s$^{-1}$. The errors in the
fluxes correspond to the fitting errors. Note that the absolute flux
calibration is estimated to be accurate to within 20\%. $^{(}$*$^{)}$
H$\beta$ is not detected in the spectrum of the nucleus and for the
northern region, it is outside the wavelength range observed, we therefore
assume H$\alpha$/H$\beta$=3.1 \cite{osterbrock89} in both cases.}
\label{tabcomaflux}
\vspace*{0.25cm}
\end{table*}

Fig.~\ref{comao3ha} shows the colour map of the [OIII]5007/H$\alpha$ line
ratio for the central region of 3C277.3, with the 1.4 GHz radio contours
superimposed.  As discussed earlier, the emission-line region coincident
with the radio knot to the south of the nucleus has higher ionization than
other regions of the galaxy, including the nucleus, which is in agreement
with previous studies by \scite{miley81} and \scite{vanbreugel85}. Its
ionization state is found to vary in the range 1.5 $\lesssim$
[OIII]/H$\alpha \lesssim$ 3.0, with the highest values corresponding to
the regions closest to the radio knots. On the other hand, the faint
emission-line region to the east of the nucleus presents lower ionization
than the nucleus, with [OIII]/H$\alpha~\sim$~0.5.

Unfortunately, the ionization state of the northern region of 3C277.3
could not be studied because of the narrower spectral range observed and
the low S/N of the [NII]6583 emission line.   

\subsubsection{Emission-line spectra}

Integrated 1-D spectra for the different emission-line regions detected in
the INTEGRAL images of 3C277.3 are presented in Fig.~\ref{speccoma} and
\ref{specNcoma} (see the captions of the figures for details of the
apertures).

The spectrum of the radio jet knot [Fig.~\ref{speccoma} (middle)] clearly
shows the narrow emission lines of this region, a sign of undisturbed
line-emitting gas, which is surprising if a strong jet-cloud interaction
is taking place in this region.  Note also the strength of the [OIII]5007
line relative to the Balmer lines in this region, indicating the high
ionization state of the gas, not expected if the emission-line gas was
ionized by jet-induced shocks.

\begin{figure}
\centerline{\psfig{figure=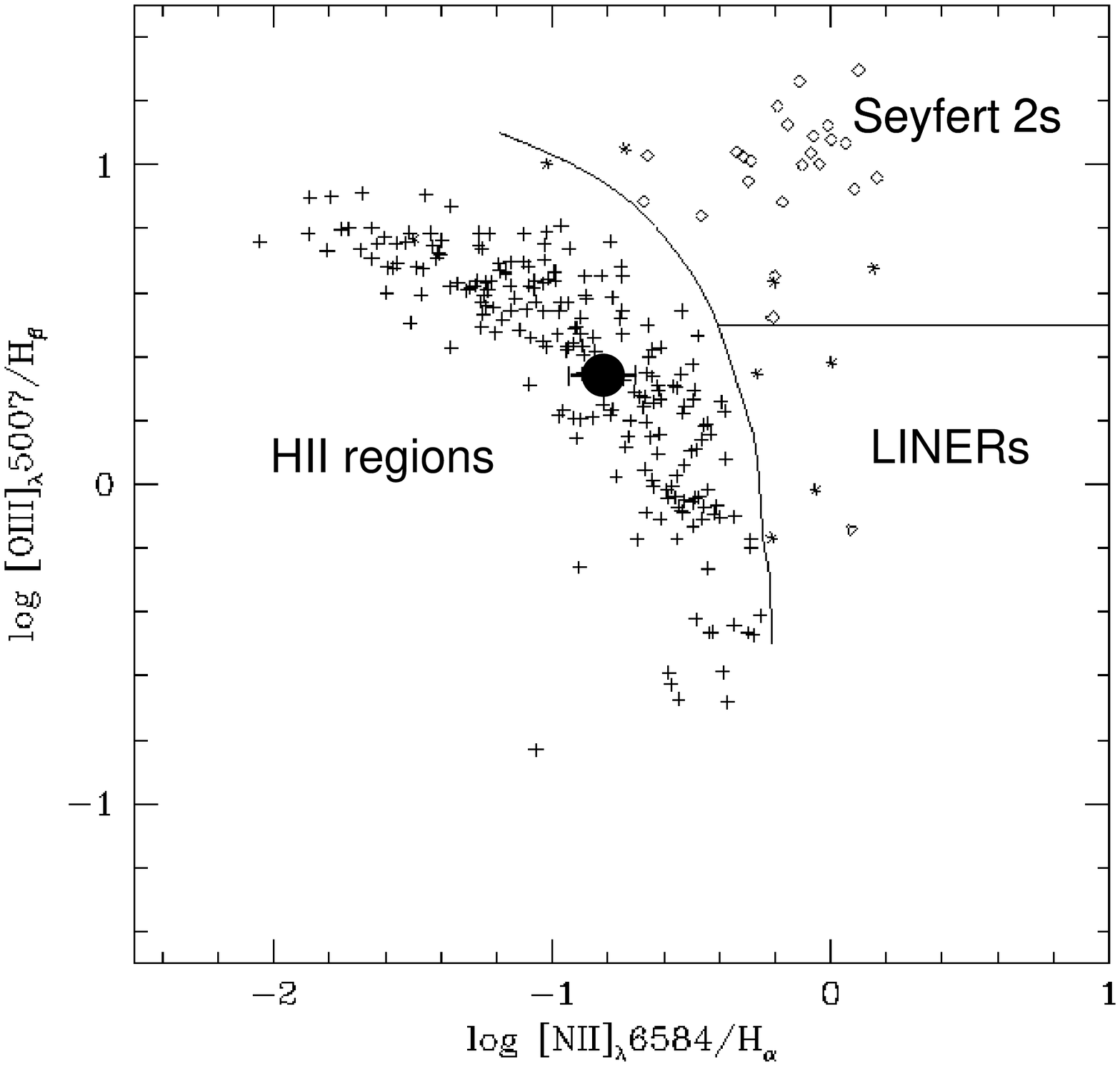,width=8cm}}
\caption[]{Diagnostic diagram taken from \scite{dessauges2000} with our
data of the eastern filament of 3C277.3 superimposed (black filled
circle). The solid lines correspond to empirical separations between
narrow-line AGNs and HII galaxies \cite{veilleux87}, and between Seyfert 2
galaxies and LINERs \cite{filippenko92b}. The crosses correspond to HII
galaxies, the diamonds to Seyfert 2 galaxies, the triangles to LINERs and
the stars to "intermediate" galaxies. Note that the eastern filament of
3C277.3 falls on the region occupied by the HII galaxies. }
\label{diagcoma}
\end{figure}

The spectrum of the eastern filament [Fig.~\ref{speccoma} (bottom)] shows
narrow emission lines and strong Balmer lines relative to the
high-ionization [OIII]5007 line, indicating that this filament is probably
an HII region. This can be clearly seen in Fig.~\ref{diagcoma}, which
shows a diagnostic diagram taken from \scite{dessauges2000}, with the line
ratio measurement of the eastern filament (black filled circle)
superimposed.  This region falls in the middle of the HII galaxies
region, far away from that of the AGNs, demonstrating that there is
ongoing star formation in the immediate environment of 3C277.3.

Fig.~\ref{specNcoma} shows the H$\alpha$ and [NII] profiles for the region
with the broadest lines (solid line) within the northern EELR of
3C277.3. Superimposed is the spectrum of the northernmost region (dashed
line), which is located beyond the radio hotspot. It is clear from this
figure that there is a marked difference in the linewidths between the two
regions. This indicates that the gas lying beyond the hotspot has not been
disturbed yet by the passage of the radio jet.

Table~\ref{tabcomaflux} lists both the absolute fluxes of the observed
emission lines and the fluxes normalised to the corresponding H$\beta$
flux for each region in 3C277.3. For the radio knot and the eastern
filament, the fluxes have not been corrected for intrinsic reddening,
given that the intensities of H$\gamma$ and H$\alpha$ relative to H$\beta$
are consistent, within errors, with the expected values for Case B
recombination \cite{osterbrock89}, suggesting that the intrinsic reddening
in these regions is not significant. For the nucleus, the H$\beta$ line
could not be detected, and therefore no estimate for the reddening in the
nucleus could be derived from our data. Thus, H$\alpha$/H$\beta$=3.1 was
assumed \cite{osterbrock89}. In the case of the northern region of
3C277.3, since H$\beta$ was outside the wavelength range observed, again
H$\alpha$/H$\beta$=3.1 was assumed.
 
\subsection{3C171}
\label{res:171}

\subsubsection{Emission-line structure}

\begin{figure}
\centerline{\psfig{figure=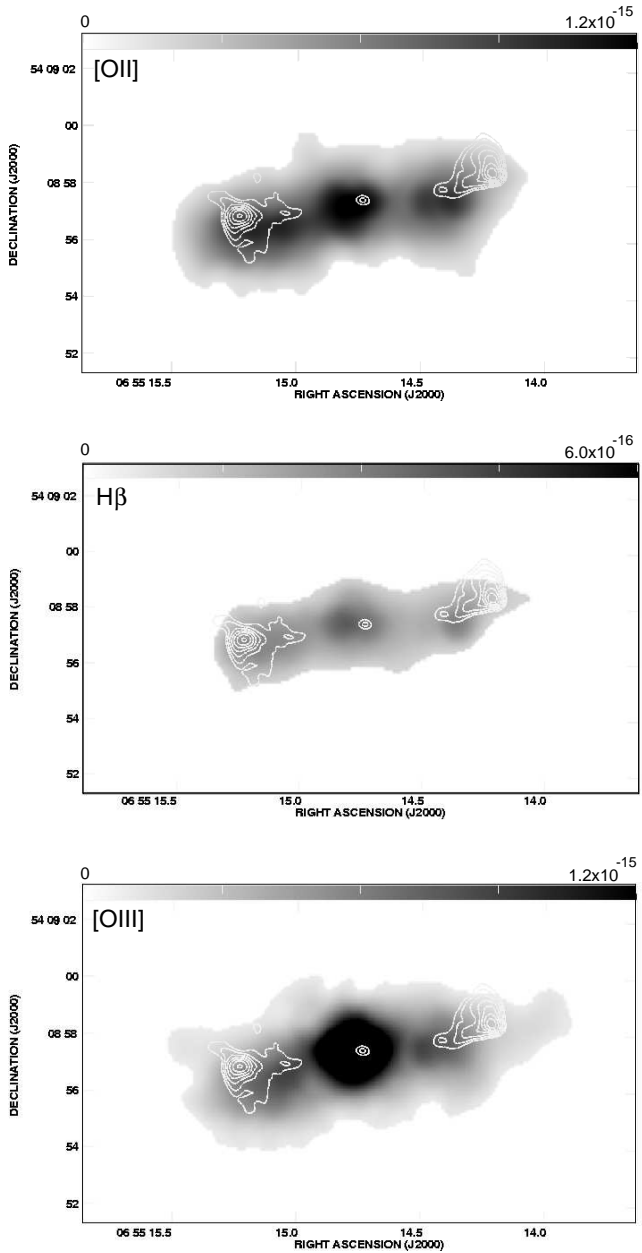,width=8.6cm}}
\caption[]{INTEGRAL grey-scale images of 3C171 in the [OII]3727 (top),
H$\beta$ (middle) and [OIII]5007 (bottom) emission lines, with the
continuum subtracted,  overlaid with the contours of the 8 GHz radio map
of \scite{hardcastle97}. The grey scale values are in units of erg
cm$^{-2}$ s$^{-1}$ arcsec$^{-2}$, and range from 0 to 1.2$\times$10$^{-15}$
for [OII] and [OIII], and from 0 to 6.0$\times$10$^{-16}$ for H$\beta$. The
values of the radio contours are 0.9 mJy beam$^{-1}$ $\times$ (-1, 1, 2,
4, 8, 16, 32, 64). }
\label{171o2hbo3}
\end{figure}

Fig.~\ref{171o2hbo3} presents the continuum-subtracted [OII]3727, H$\beta$
and [OIII]5007 emission-line INTEGRAL images for 3C171. The contours of
the 8 GHz radio emission (from \pcite{hardcastle97}) are superimposed. To
align the radio and optical maps, the radio core was assumed to be
coincident with the optical continuum centroid of the galaxy.

It can be seen that the emission-line structure of 3C171 is elongated and
extends about 14 arcsec (67 kpc) along the radio axis of this
source. Further, the optical line emission is clearly spatially associated
with the radio emission: bright regions of [OII], H$\beta$ and [OIII]
emission are detected on either side of the nuclear region, close to the
radio lobes. Note also that the radio hotspots lie outside the brightest
regions on either side of the nucleus. It can also be noticed that in both
eastern and western enhanced regions the [OII] and H$\beta$ emissions
appear to be brighter relative to the nucleus than the emission in [OIII],
which indicates that the ionization state of both eastern and western EELR
is lower than that of the nucleus.  From the images in
Fig.~\ref{171o2hbo3} it is also apparent that the line emission centroids
are significantly displaced towards the east of the radio core (and
therefore the continuum centroid). This is more noticeable in the [OII]
and H$\beta$ emission, than in [OIII].

\subsubsection{Emission-line kinematics}

\begin{figure*}
\centerline{\psfig{figure=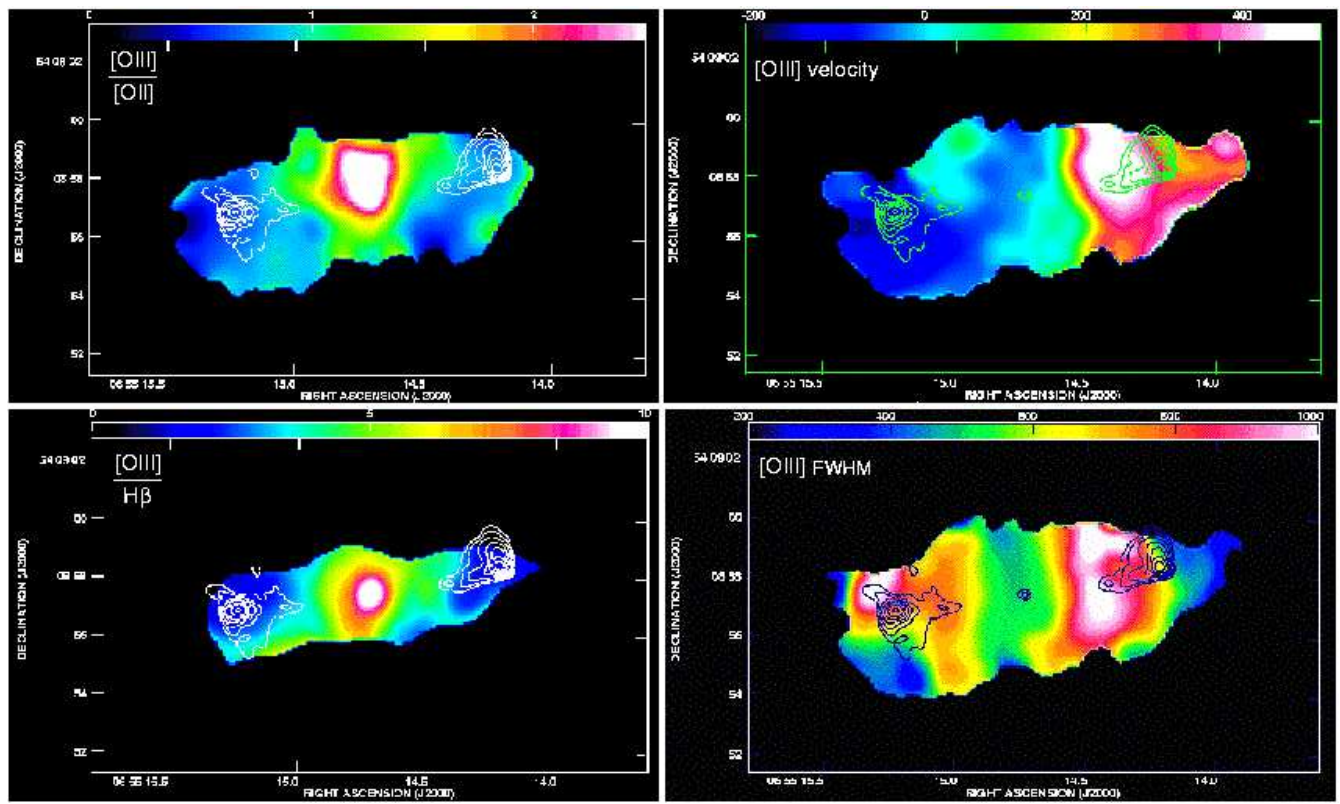,angle=-90,width=17.75cm}}
\vspace{-0.1cm}
\caption[]{Colour maps of 3C171 showing the [OIII] velocity (top right),
[OIII] instrumentally-corrected linewidth (bottom right) and the
[OIII]/[OII] (top left) and [OIII]/H$\beta$ (bottom left) line ratios. The
contours of the radio emission at 8 GHz (from \pcite{hardcastle97}) are
superimposed. Radio contour levels are as in Fig.~\ref{171o2hbo3}.}
\label{171ratvelwid}
\end{figure*}

The maps of the velocity field and (instrumentally-corrected) linewidth for
the emission-line gas in 3C171 are presented in Fig.~\ref{171ratvelwid}
(right). Contours of the 8 GHz radio emission (from \pcite{hardcastle97}) are
superimposed.  

The velocity field appears to be complex [Fig.~\ref{171ratvelwid} (top
right)]. It shows a gradient increasing from east to west, with an overall
velocity amplitude of $\Delta v \sim$ 600 \kms.  The eastern extreme of the
emission-line structure is blueshifted with a velocity shift of $\sim$
--200 \kms. The relative velocity of the gas rises smoothly from
there up to a location between $\sim$ 1.5 and 2.5 arcsec (7 - 12 kpc) west
from the nucleus, where there is a sudden increase in the velocity from
$\sim$ +200 to +500 \kms. The velocity then decreases slightly towards the
western extreme of the emission-line structure to a value of $\sim$ +300
\kms.

Fig.~\ref{171ratvelwid} (bottom right) shows the variation of the [OIII]
linewidth over the emission-line structure of 3C171. In the nuclear region
the linewidths are $\sim$ 550 \kms \ (FWHM). Then they rise towards the
east and west of the nucleus reaching values of $\sim$ 1000 -- 1100 \kms \
(FWHM) in the regions just behind the radio lobes, suggesting that it is
emission from the shocked gas cooling behind the shock front.  Beyond the
radio hotspots the linewidths are observed to decrease to values of $\sim$
400 \kms \ (FWHM). The close spatial association between the regions with
the broadest emission lines and the radio structure in 3C171 suggests that
the ambient gas at these locations has been perturbed by interactions with
the radio source. The regions with the broadest emission lines extend over
$\sim$ 4 arcsec (20 kpc) in the direction perpendicular to the radio axis,
indicating that their extreme kinematics are the result of the lateral
expansion of the radio cocoon. The effects of the jet-cloud interaction
are seen across the entire jet cocoon, not just the narrow region along
the radio jet.

\begin{figure}
\centerline{\psfig{figure=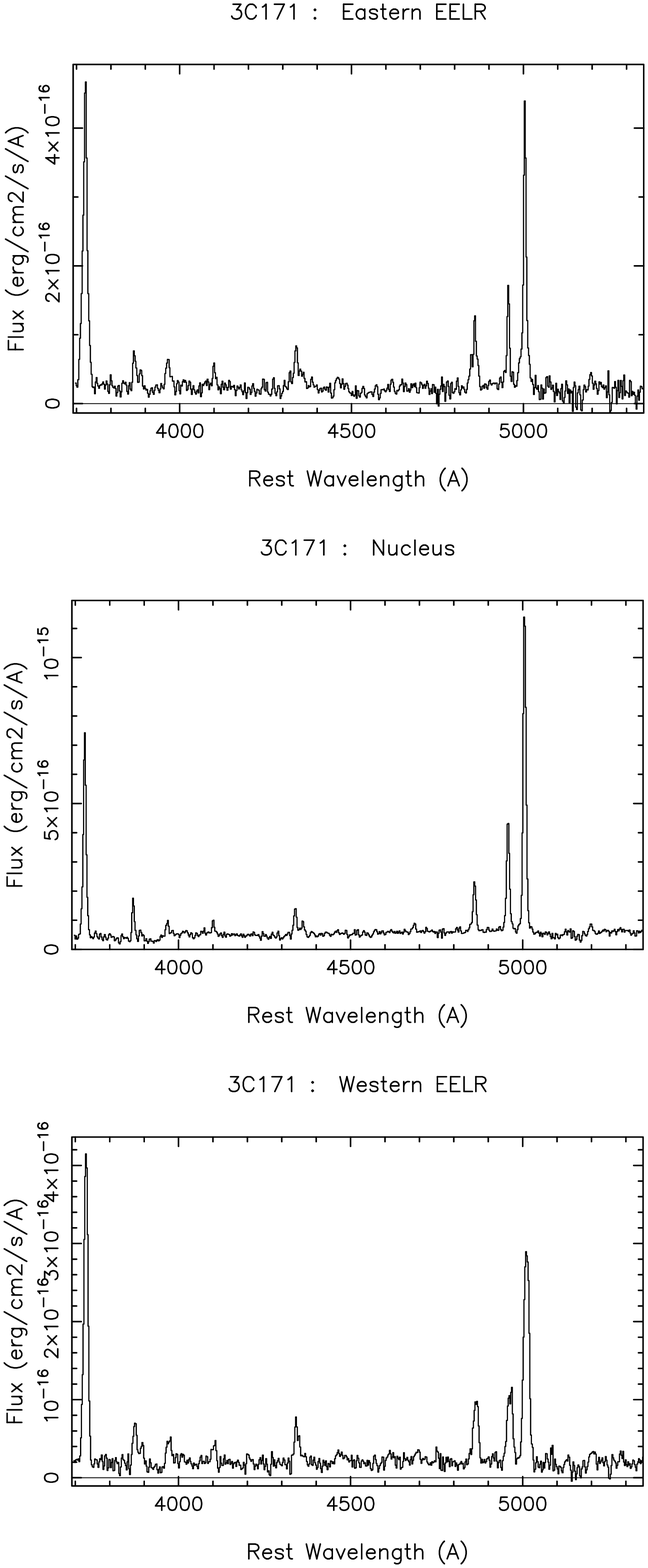,width=8.5cm}}
\caption[]{Integrated spectra of the different regions in 3C171:
Eastern EELR (top: 13.7 arcsec$^2$ aperture centred at 4 arcsec east from
the continuum centroid), nucleus (middle: 12.2 arcsec$^2$ aperture) and
western EELR (bottom: 11.4 arcsec$^2$ aperture centred at 3.1 arcsec west
form the continuum centroid).} 
\label{spec171}
\end{figure}

\subsubsection{Emission-line ionization state}

Maps of the [OIII]/[OII] and [OIII]/H$\beta$ line ratios for 3C171 are
shown in Fig.~\ref{171ratvelwid} (left).
These images clearly show that the nuclear region has the highest
ionization state ([OIII]/[OII] $\simeq$ 3 and
[OIII]/H$\beta$~$\simeq$~10). As the radius increases, the ionization
state of the gas is observed to decrease reaching minimum values of
[OIII]/[OII]~$\simeq$~0.4 -- 0.7 and [OIII]/H$\beta$ $\simeq$ 2 at the
location of the radio hotspots.
It is noticeable, from the [OIII]/[OII] map (top left), that beyond the
western radio hotspot the ionization state of the gas seems to start
increasing again towards the extreme of the emission-line structure. This,
however, is not observed so clearly beyond the eastern radio lobe.

\subsubsection{Emission-line spectra}

Integrated 1-D spectra for the nuclear region, the western and eastern
EELR of 3C171 are presented in Fig.~\ref{spec171} (see the caption of the
figure for details of the apertures). The relative intensities of the
[OII]$\lambda3727$ and [OIII]$\lambda5007$ lines in the three
emission-line regions, clearly demonstrate the lower ionization state of
the eastern and western EELR relative to the nucleus.

Both the absolute fluxes of the observed emission lines and the normalised
fluxes to the corresponding H$\beta$ flux for each region in 3C171 are
listed in Table~\ref{tab171flux}. Given that the observed
H$\delta$/H$\beta$ and H$\gamma$/H$\beta$ line ratios are consistent
(within errors) with the theoretical Case B recombination
\cite{osterbrock89}, no corrections for intrinsic reddening have been made
to the line fluxes.

\subsection{3C265}

\subsubsection{Emission-line structure}

Fig.~\ref{265o2} shows the continuum-subtracted [OII]3727 emission-line
INTEGRAL image for 3C265. By contrast to 3C277.3 and 3C171, the radio
emission in 3C265 is much more extended than the emission-line structures
(see radio map in \pcite{fernini93}), and so in the figures only
the radio axis is indicated (black solid line).  

This image shows that, in contrast to 3C171 and other high-redshift radio
galaxies, the emission-line structure of 3C265 within $\sim$10 arcsec (48
kpc) of the nucleus does not align along the radio axis\footnote{Note,
however, that the emission-line structure of 3C265 at radii $\gtrsim$ 10
arcsec (82 kpc) is observed to be closely aligned along the radio axis on
the eastern side of the nucleus (e.g.  Sol{\'o}rzano-I{\~n}arrea et
al. 2002; cf. Fig~\ref{cones265}).}. It is in fact elongated and
misaligned by $\sim$ 35$^{\circ}$ with respect to the radio axis, which is
still consistent with the large ($\sim$60$^{\circ}$) half opening angle of
the ionization bicone detected in deep narrow-band images of this source
(see Fig.~\ref{cones265}; Sol{\'o}rzano-I{\~n}arrea et al. 2002).
  
\subsubsection{Emission-line kinematics}

The velocity field and (instrumentally-corrected) linewidth maps for
3C265 are presented in Fig.~\ref{265velwidth}. The white solid line
indicates the radio axis.

The velocity of the emission-line gas (left) appears to follow a smooth
gradient, with the exception of a region located in the south-western
extreme of the emission-line structure at $\sim$ 2 arcsec from the
nucleus. The gas in this region presents a sharp change in the velocity of
approximately 300 \kms \ (from --50 to +250 \kms) along a projected
distance of only $\sim$ 2 arcsec (16.5 kpc).

Apart from this region, the rest of the emission-line gas presents an
ordered velocity profile with an overall amplitude of $\Delta v \lesssim$
600 \kms, which is consistent with gravitational motions in the galaxy
(\pcite{tadhunter89b,baum90}). The gas at the south-eastern extreme of the
emission-line structure is on average blueshifted with a velocity of
$\sim$ --150 \kms, and the north-western blob is redshifted with a
velocity of $\sim$ +400 \kms.
 
The variation of the [OII] linewidths over the emission-line structure of
3C265 is shown in Fig.~\ref{265velwidth} (right). The linewidths are in
the range $\sim$ 600 -- 700 \kms \ (FWHM) over most of the nebula, with the
exception of two regions where significant increases in the widths of the
lines are observed.

The region located about 2 arcsec (16.5 kpc) east from the nucleus, and
extending over 4.5 arcsec (37 kpc) in the north-south direction, presents
linewidths in the range $\sim$ 950 -- 1150 \kms \ (FWHM). This region is
situated on the passage of the eastern radio jet, which suggests that the
broadening of the emission lines in this region might be caused by
jet-cloud interactions. It should be noted that at this location, $\sim$
2.5 arcsec east from the nucleus of 3C265, a strong line splitting
($\Delta v \sim$ 1000 \kms) has been detected in the [OIII] emission line
(\pcite{tadhunter91}; Sol{\'o}rzano-I{\~n}arrea et al. 2002), the most
likely origin of which is jet-induced shocks. Note, however, that this
broad-line region has a much larger extent in the north-south direction
that the high-velocity knot.

The region situated about 2 arcsec to the south-west of the nucleus,
extends approximately over 3.5 arcsec in the north-south direction, and
presents linewidths in the range $\sim$ 950 -- 1350 \kms \ (FWHM). This
region, however, is offset from the passage of the western radio jet.

\subsubsection{Emission-line spectra}

\begin{table*}
\small
\begin{center}
\begin{tabular}{lccccccccc}\hline \\
\ \ {\bf \large 3C171}& \ &\multicolumn{2}{c}{\bf Eastern EELR}& \
 &\multicolumn{2}{c}{\bf Nucleus}& \ &\multicolumn{2}{c}{\bf Western EELR} \\
 & & Flux abs. & Flux ratio & & Flux abs. & Flux ratio & &Flux abs. & Flux
 ratio \\ 
{\bf Line} & & $/ 10^{-15}$ & to H$\beta$& & $/ 10^{-15}$ & to H$\beta$& & $/
 10^{-15}$ & to H$\beta$ \\ \\ \hline \\

$[$OII$]\lambda$3727 & & 
8.23$\pm$0.17 & 3.99$\pm$0.20 & & 8.71$\pm$0.13 & 3.72$\pm$0.18 & & 7.28$\pm$0.08 & 3.99$\pm$0.22 \\

$[$NeIII$]\lambda$3869 & & 
0.71$\pm$0.12 & 0.34$\pm$0.06 & & 1.29$\pm$0.10& 0.55$\pm$0.05 & & 0.94$\pm$0.08 & 0.52$\pm$0.05 \\

H8+He\hspace{0.4ex}I$\lambda$3886 & & 
0.26$\pm$0.10 & 0.13$\pm$0.05 & & 0.21$\pm$0.08 & 0.09$\pm$0.04 & & 0.36$\pm$0.07 & 0.20$\pm$0.04 \\

$[$NeIII$]\lambda$3967 & & 
0.63$\pm$0.13 & 0.31$\pm$0.06 & & 0.61$\pm$0.10& 0.26$\pm$0.04 & & 0.63$\pm$0.08 & 0.35$\pm$0.05 \\

H$\delta$ & & 
0.41$\pm$0.07 & 0.20$\pm$0.04 & & 0.53$\pm$0.07 &0.23$\pm$0.03 & & 0.45$\pm$0.07 & 0.24$\pm$0.04 \\

H$\gamma$ & & 
1.07$\pm$0.13 & 0.52$\pm$0.06 & & 1.17$\pm$0.07 &0.50$\pm$0.04 & & 1.01$\pm$0.07 & 0.55$\pm$0.05\\

$[$OIII$]\lambda$4363 & & 
0.38$\pm$0.10 & 0.19$\pm$0.05 & & 0.67$\pm$0.08& 0.29$\pm$0.04 & & 0.18$\pm$0.07 & 0.10$\pm$0.03\\

HeII\hspace{0.4ex}$\lambda$4686  & & 
--- & --- & & 0.36$\pm$0.10 & 0.14$\pm$0.04 & & 0.36$\pm$0.11 & 0.20$\pm$0.06\\

H$\beta$  & & 
2.06$\pm$0.10 & 1.00$\pm$0.05 & & 2.34$\pm$0.10 & 1.00$\pm$0.04 & & 1.82$\pm$0.10 & 1.00$\pm$0.05\\

$[$OIII$]\lambda$4959 & & 
1.98$\pm$0.10 & 0.94$\pm$0.07 & & 5.00$\pm$0.08 & 2.13$\pm$0.10 & & 2.24$\pm$0.08 & 1.23$\pm$0.08\\

$[$OIII$]\lambda$5007 & & 
5.81$\pm$0.13 & 2.81$\pm$0.15 & & 14.45$\pm$0.10 & 6.17$\pm$0.28 & & 6.63$\pm$0.10 & 3.63$\pm$0.20\\

$[$NI$]\lambda$5199 & & 
0.24$\pm$0.08 & 0.13$\pm$0.04 & & 0.33$\pm$0.10 &0.14$\pm$0.04 & & 0.31$\pm$0.08 & 0.17$\pm$0.05\\ \\ \hline

\end{tabular}
\end{center}
\vspace{-0.2cm}
\caption[]{Emission-line integrated fluxes for the different regions in
3C171. The details of the apertures are in the caption of
Fig.~\ref{spec171}.  The absolute fluxes are in units of erg cm$^{-2}$
s$^{-1}$. The errors in the fluxes correspond to the fitting errors. Note
that the absolute flux calibration is estimated to be accurate to within
20\%.}
\label{tab171flux}
\vspace*{0.25cm}
\end{table*}

\begin{figure*}
\centerline{\psfig{figure=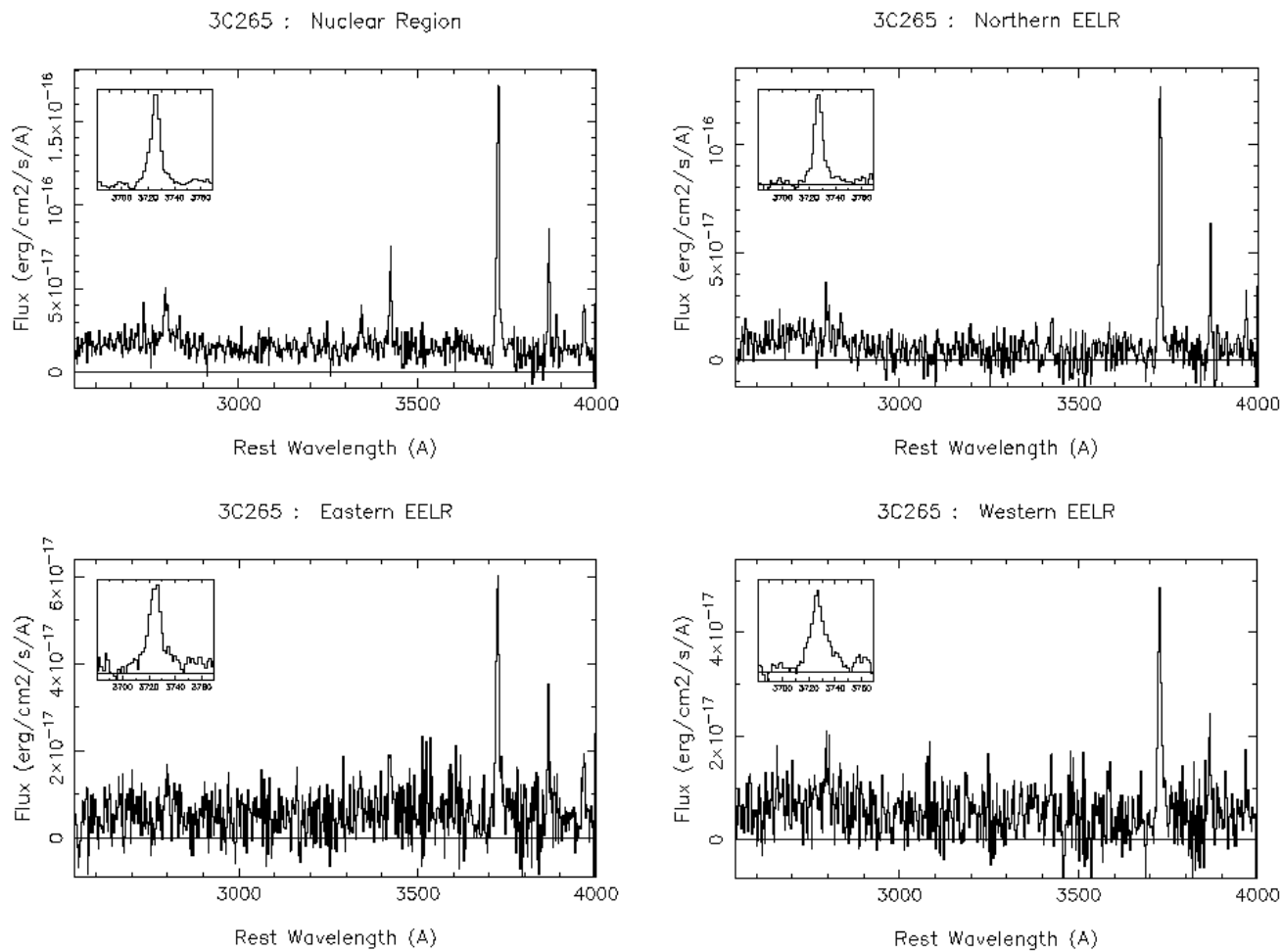,width=17.6cm}}
\vspace{-0.15cm}
\caption[]{Integrated spectra for the different regions in 3C265: Nuclear
region (top left: 6.1 arcsec$^2$ aperture), northern EELR (top right: 7.6
arcsec$^2$ aperture centred at 4.2 arcsec north-west of the continuum
centroid), eastern EELR (bottom left: 4.6 arcsec$^2$ aperture centred at
1.5 arcsec east of the continuum centroid) and western EELR (bottom right:
4.6 arcsec$^2$ aperture centred at 1.9 arcsec west of the continuum
centroid). The insets show the [OII]$\lambda$3727 region of the
spectrum. }
\label{spec265}
\end{figure*}

Fig.~\ref{spec265} presents the integrated 1-D spectra for different
emission-line regions in 3C265: nucleus (top left), northern EELR (top
right), eastern EELR (bottom left) and western EELR (bottom right). 
The details of the apertures are given in the caption of the figure.

Both the absolute fluxes of the observed emission lines and the fluxes
normalised to the corresponding [OII]$\lambda$3727 (since H$\beta$ was not
covered in the observed spectral range) are listed for each region in
Table~\ref{tab265flux}.

\section{Discussion}

\subsection{Jet-cloud interactions and the role of the radio jet cocoon}

As discussed in the introduction, the three galaxies studied here were
previously known to be undergoing jet-cloud interactions at some level.
Evidence for this has been mainly obtained from long-slit spectroscopic
studies. The new INTEGRAL observations allow an investigation of how the
jet-cloud interactions evolve when moving away from the radio axis, and
also a study of any changes in the dominant mechanism of the emission-line
gas over the entire nebula of the galaxies. 

Amongst the three galaxies studied in this paper, 3C171 shows most clearly
that the properties of the emission-line gas in this galaxy are defined by
interactions between the radio source and the ambient gas.  Several
results provide evidence for this.

\begin{figure}
\centerline{ \psfig{figure=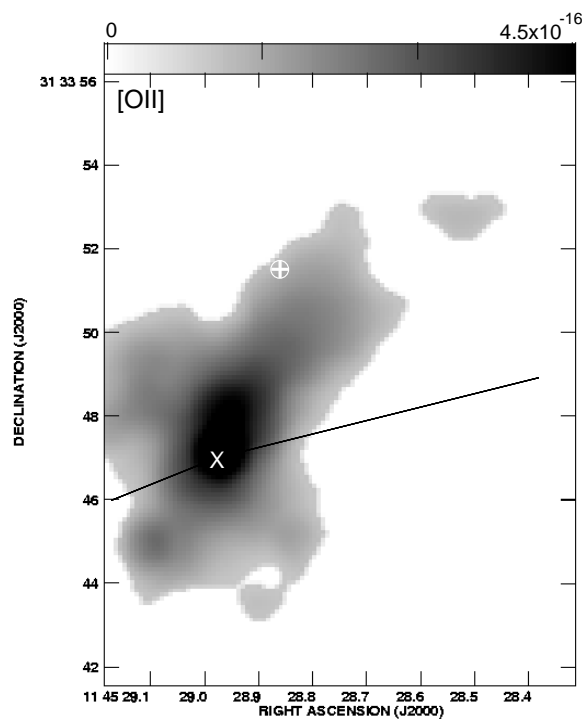,width=7.8cm}}
\caption[]{INTEGRAL grey-scale image of 3C265 in the [OII]3727 emission
line, with the continuum subtracted. The continuum centroid is indicated
by a white `x', and the solid line represents the radio axis. The white
`+' in a circle indicates the position of the nearest companion
galaxy. The grey scale values range from 0 to 4.5$\times$10$^{-16}$ erg
cm$^{-2}$ s$^{-1}$ arcsec$^{-2}$.}
\label{265o2}
\end{figure}

Firstly, the enhanced brightness of the line emission along the radio axis
observed in the [OII], H$\beta$ and [OIII] intensity maps
(Fig.~\ref{171o2hbo3}) indicates that either jet-induced shocks are acting
as an effective local ionization source, or that as the gas is swept up
and compressed by the passage of the jet, the shocks increase the
efficiency with which the central AGN photoionizes the ambient gas (see
also \pcite{clark98}).

\begin{figure*}
\centerline{\psfig{figure=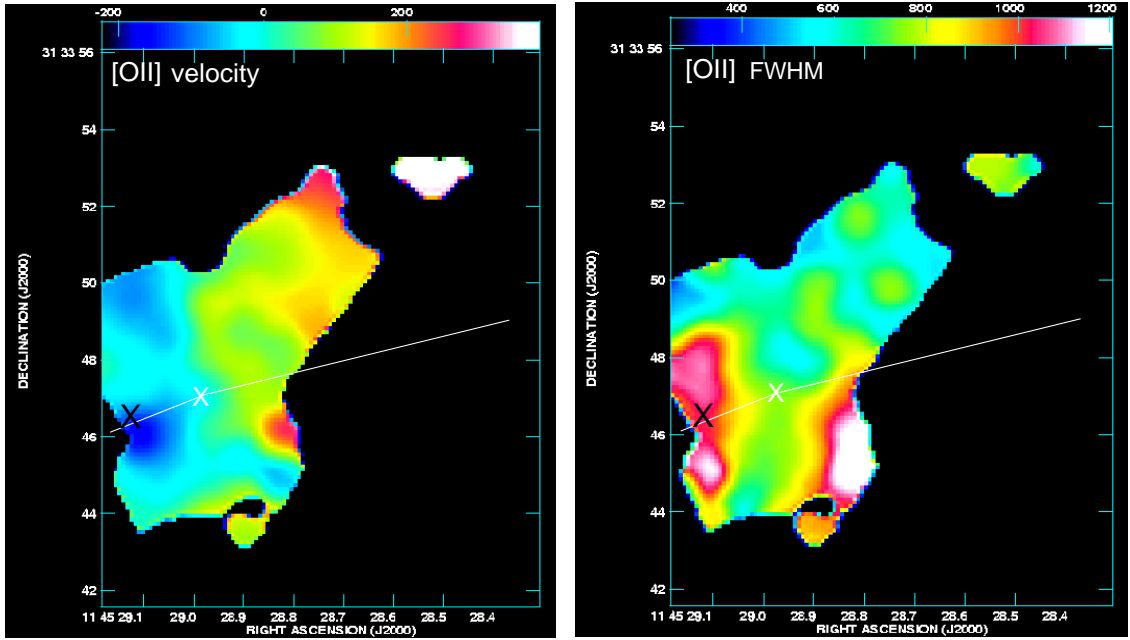,width=15cm}}
\vspace{-0.15cm}
\caption[]{[OII] velocity (left) and instrumentally-corrected linewidth
(right) colour maps for 3C265. The continuum centroid and the radio axis
are represented by the white `x' and solid line, respectively. The black
`x' indicates the position of the high-velocity cloud
(Sol\'{o}rzano-I\~{n}arrea et al. 2002).}
\label{265velwidth}
\end{figure*}

Secondly, regions with a sudden increase in the linewidth are observed
close to the location of the hotspots (Fig.~\ref{171ratvelwid}, bottom
right), extending in the direction perpendicular to the radio axis, up to
the edge of the observed emission-line structure. This rise in the FWHM
reflects the existence of multiple kinematic components, including line
splittings and underlying broad components.  The nature of the different
components in the kinematic structure of the emission-line gas along the
radio axis in 3C171 has been studied in detail by \scite{clark98}. Such
extreme kinematics, which cannot be explained by gravitational motions,
indicate that the gas has been highly perturbed, and must be the result of
acceleration induced by the passage of the jet through the ambient
gas. Now the INTEGRAL results show that these complex kinematics also
extend perpendicular to the radio axis, indicating that
the effects of the radio jets are also felt far from the radio axis.

Thirdly, the regions with the lowest ionization state are coincident with
the location of the radio lobes (Fig.~\ref{171ratvelwid}, left). This can
be explained either by local shock-ionization\footnote{Shock-ionization
models predict lower ionization states than the photoionization
predictions.}, or by a shock-induced increase of the local density
followed by AGN-photoionization\footnote{Note that the ionization state of
photoionized gas is inversely proportional to the density of the gas.}.
Detailed examination of the emission-line ratios along the radio axis of
3C171 by \scite{clark98} shows, however, that AGN-photoionization models
fail to explain certain features measured in the EELR of 3C171, such as
the low HeII/H$\beta$ ratios and the high [OIII] electron temperatures,
both of which can be accounted for in the shock-ionization predictions
(e.g. \pcite{dopita95}, 1996)\nocite{dopita96}.

In conclusion, these INTEGRAL images provide further strong evidence
that the EELR properties in 3C171 are to a large extent defined by
interactions with the radio jet. In addition, these images show for the first
time that the regions with the broadest emission lines and the lowest
ionization state are found not only near the radio axis, but also
extending in the direction perpendicular to this. This indicates that
the lateral expansion of the radio source cocoon has a significant 
effect on the kinematics and ionization of the ambient gas.

\medskip

Considering now the results obtained for 3C277.3, a few similarities can
be drawn between the observed emission-line properties of 3C277.3 and
those of 3C171.  In the first place, regions of enhanced emission-line
luminosity are observed coincident with the northern radio hotspot
(Fig.\ref{comaNh}) and with the radio jet knot to the south of the nucleus
(Fig.\ref{comanucfl}), suggesting that they are the result of strong
interactions between the radio source and the ambient gas.  Moreover, the
observed increase in the widths of the emission lines in the region of the
northern hotspot, together with the sudden decrease of the linewidths
observed just beyond the radio hotspot (Fig.\ref{comavelwidth}, bottom
right), is further clear evidence that the emission-line gas at this
location is being disturbed by interactions with the radio structure.

On the other hand, the region coincident with the radio jet knot, which is
clearly undergoing a jet-cloud interaction, does not present any disturbed
emission-line kinematics. This is not expected if the enhanced line
emission in this region is the result of a strong collision between the
radio jet and a massive cloud that causes the radio jet to be deflected
\cite{vanbreugel85}. This puzzling result will be discussed in
Section~\ref{diss:comaknot}.

\medskip

Both 3C171 and 3C277.3 have radio source sizes of comparable physical
extents to their emission-line structures, and thus it is not surprising
that the radio structure is interacting with the host galaxy in both
cases.  By contrast, the radio size of 3C265 is much larger than the
emission-line nebula, and yet this source still presents signs of strong
jet-cloud interactions in the emission-line kinematics (see also
\pcite{tadhunter91}) \footnote{Note that 3C265 is not the only large radio
source in which kinematic evidence for jet-cloud interactions has been
found \cite{carmen01}.}.  On the other hand, the emission-line gas of
3C265 is known to be mainly photoionized by the central AGN. Moreover, the
ionization structure within $\sim$10 arcsec of the nucleus appears to be
dominated by the predicted illumination cones (see Fig.~\ref{cones265}).

The INTEGRAL images of 3C265 reveal two kinematically disturbed regions,
with very broad (FWHM $\sim$ 950 --- 1300 \kms) emission lines, at about 2
arcsec east and west of the nucleus (Fig.~\ref{265velwidth}, right). These
regions have a large lateral extent ($\sim$ 30 --- 38 kpc), showing a
similar effect to that observed in 3C171. Thus, 3C265 is another case in
which the lateral expansion of the radio cocoon around the jet may have an
important effect on the kinematics of the emission-line gas.
Alternatively, the disturbed kinematics of these two regions in 3C265
could also be the result of strong outflows driven by the central AGN in
the ionization cones. If this is the case, the broadening of the lines
would be a consequence of seeing both the near and far sides of the cone
simultaneously. A sign for this could be the sharp velocity change
observed in the western broad-line region, where we observe blueshifted
gas moving at approximately --50 \kms \ and redshifted gas at +250 \kms \
both with respect to the nucleus (Fig.~\ref{265velwidth}, left). It should
be noted that evidence for AGN-induced outflows has been previously found
at low redshifts for another powerful radio galaxy: Cygnus A
\cite{tadhunter99}. Moreover, deep narrow-band images of 3C265
(Sol{\'o}rzano-I{\~n}arrea et al. 2002) show the emission-line structure
with an apparent ``hollowing out'' of the cones, which is entirely
consistent with outflows driven by the AGN.

\medskip

\begin{table*}
\footnotesize
\begin{center}
\hspace*{-0.557cm}
\begin{tabular}{lcccccccccccc}\hline \\
\ \ {\bf \large 3C265}& &\multicolumn{2}{c}{\bf Eastern EELR}& \
 &\multicolumn{2}{c}{\bf Nuclear Region}& \ &\multicolumn{2}{c}{\bf
 Western EELR}& \ &\multicolumn{2}{c}{\bf Northern EELR} \\
 & & Flux abs. & Flux ratio & & Flux abs. & Flux ratio & &Flux abs. & Flux
 ratio & & Flux abs. & Flux ratio   \\ 
{\bf Line} & & $/ 10^{-16}$ & to $[$OII$]$& & $/ 10^{-16}$ & to $[$OII$]$ & & $/
 10^{-16}$ & to $[$OII$]$& & $/ 10^{-16}$ & to $[$OII$]$ \\ \\ \hline \\

HeII\hspace{0.4ex}$\lambda$2733 & &
 --- & --- & & 1.72$\pm$0.38 & 0.06$\pm$0.02 & & --- & --- & & --- & --- \\

MgII\hspace{0.4ex}$\lambda$2798 & & 
1.93$\pm$0.45 & 0.17$\pm$0.04 & & 8.10$\pm$0.84 & 0.29$\pm$0.04 & & 2.54$\pm$0.51 & 0.27$\pm$0.06 & & 3.74$\pm$0.59 & 0.21$\pm$0.03 \\ 

$[$OIII$]\lambda$2837 & &
 --- & --- & & 1.95$\pm$0.61 & 0.07$\pm$0.02 & & 0.87$\pm$0.31 & 0.09$\pm$0.03 & & 1.56$\pm$0.46 & 0.09$\pm$0.02 \\

$[$NeV$]\lambda$3346 & &
1.22$\pm$0.41 & 0.11$\pm$0.04 & & 3.02$\pm$0.46 & 0.12$\pm$0.02 & & 0.66$\pm$0.30 & 0.07$\pm$0.03 & & 1.19$\pm$0.43 & 0.07$\pm$0.02 \\

$[$NeV$]\lambda$3426 & & 
2.49$\pm$0.51 & 0.22$\pm$0.05 & & 7.67$\pm$0.54 & 0.28$\pm$0.03 & & 1.27$\pm$0.31 & 0.13$\pm$0.03 & & 1.93$\pm$0.45 & 0.11$\pm$0.02 \\ 

$[$OII$]\lambda$3727  & &
11.37$\pm$0.79 & 1.00$\pm$0.07 & & 27.59$\pm$1.98 & 1.00$\pm$0.07 & & 9.50$\pm$0.73 & 1.00$\pm$0.08 & & 18.15$\pm$0.76 & 1.00$\pm$0.04 \\

$[$NeIII$]\lambda$3869 & &
 3.78$\pm$0.64 & 0.33$\pm$0.06 & & 9.39$\pm$0.64 & 0.34$\pm$0.03 & & 1.80$\pm$0.41 & 0.19$\pm$0.05 & & 6.57$\pm$0.63 & 0.36$\pm$0.04 \\

H8+He\hspace{0.4ex}I$\lambda$3886 & & 
 --- & --- & & 1.77$\pm$0.53 & 0.06$\pm$0.02 & & --- & --- & & 9.94$\pm$0.43 & 0.05$\pm$0.02 \\ 

$[$NeIII$]\lambda$3967 & &
 2.01$\pm$0.56 & 0.18$\pm$0.05 & & 3.91$\pm$0.56 & 0.14$\pm$0.02 & & 1.30$\pm$0.38 & 0.14$\pm$0.04 & & 3.22$\pm$0.54 & 0.18$\pm$0.03 \\ \\ \hline

\end{tabular}
\end{center}
\vspace{-0.2cm}
\caption[]{Emission-line integrated fluxes for the different regions in
3C265. The details of the apertures are in the caption of
Fig.~\ref{spec265}. The absolute fluxes are in units of erg cm$^{-2}$
s$^{-1}$. The errors in the fluxes correspond to the fitting errors. Note
that the absolute flux calibration is estimated to be accurate to within
20\%. Since H$\beta$ was not covered in the observed wavelength range, the
relative fluxes are normalised to the [OII] flux instead.}
\label{tab265flux}
\vspace*{0.25cm}
\end{table*}

In summary, although it is difficult to generalize from observations of
three objects at very different redshifts, some common properties which
show evidence for interactions between the ambient gas and the radio
source are found in the EELR of all three sources, such as enhanced
emission-line luminosities and disturbed kinematics associated with the
radio structures. Perhaps one of the most interesting results obtained
from these INTEGRAL images is the fact that the lateral expansion of the
radio cocoon plays an important role in the properties of the
emission-line gas, at least for 3C171 and 3C265. In addition, the images
of 3C265 also show that AGN-induced outflows might have a significant
effect in the kinematics of the ambient gas in this galaxy.

\subsection{The nature of the emission-line gas in the radio jet knot
of 3C277.3} 
\label{diss:comaknot}

As discussed earlier, despite the strong morphological evidence for a
jet-cloud interaction taking place in the emission-line region of the
radio jet knot in 3C277.3, surprisingly no kinematic signs of such an
interaction are detected in the emission-line gas: no steep gradients in
the velocity field are observed --- the whole region is uniformly moving
at a projected velocity of approximately --200~\kms \ with respect to the
nucleus (Fig.~\ref{comavelwidth}, top left) --- and most of the region
presents narrow emission lines (Fig.~\ref{speccoma}, middle).

\medskip

We first consider whether the central AGN is energetically capable of
producing the observed emission-line luminosity in the region of the radio
jet knot. 

For a typical 3C radio source at a similar redshift of 3C277.3 (z $\sim$
0.08), the central ionizing photon production rate is $Q\sim 10^{53}$
phot~s$^{-1}$ \cite{thesismccarthy88}. The number of ionizing photons
intercepted by the emission-line knot will depend on the fraction of sky
that it covers. Assuming that it has spherical symmetry as seen from the
nucleus, we estimate that the solid angle subtended by this region from
the nucleus is $\Omega \sim$~0.9~sr. Thus, the number of ionizing photons
per unit time intercepted by the emission-line knot is $\sim$
7.1$\times$10$^{51}$ phot~s$^{-1}$.  The H$\beta$ luminosity L(H$\beta$)
that the knot would emit due to the central AGN ionization can be
estimated by using the following formula \cite{osterbrock89}:
\begin{equation}
L(H\beta) = h\nu_{H\beta}n_{e}^{2}fV\alpha_{H\beta}^{eff}
\end{equation}
\noindent
where $h$ is the Planck's constant, $\nu_{H\beta}$ is the H$\beta$
frequency, $n_{e}$ is the electron density, $V$ is the volume of the
emitting region, $f$ is the volume filling factor and
$\alpha_{H\beta}^{eff}$ is the effective recombination coefficient.
Taking into account that the number of ionizing photons intercepted by the
region is $n_{e}^{2}fV\alpha_{B}$, where $\alpha_{B}$ is the total
recombination coefficient, and assuming Case B recombination
\cite{osterbrock89}, we obtain that the H$\beta$ luminosity due to the
AGN\footnote{Any additional blazar beam is unlikely to affect the
ionization of the region significantly, given that the opening angle
subtended by this region from the nucleus is $\sim$60$^{\circ}$.}  would
be L(H$\beta$)$_{\rm AGN}$$\sim$3.4$\times$10$^{39}$ erg~s$^{-1}$. Using
the relation H$\alpha$/H$\beta$=3.1 of Case B, we find that the luminosity
of H$\alpha$ emitted by the knot due to the central AGN would be
L(H$\alpha$)$_{\rm AGN}$$\sim$1.1$\times$10$^{40}$~erg~s$^{-1}$.

The H$\alpha$ luminosity actually emitted by the knot, based on
the observed H$\alpha$ flux (see Table~\ref{tabcomaflux}), is
L(H$\alpha$)$_{\rm obs}$ $\sim$ 2.7$\times$10$^{41}$ erg~ s$^{-1}$, which
exceeds by more than an order of magnitude the expected H$\alpha$
luminosity due to the AGN.  Therefore, an extra source of ionization is
needed to produce the observed H$\alpha$ luminosity in the region of the
radio jet knot.

Jet-induced shocks are likely to be this ionizing source. To consider
whether this local mechanism is energetically viable, we first calculate
the total kinetic power of the jet $K_{jet}$ using the following formula
given in \scite{rawlings91}:

\begin{equation}
\label{eq:jet}
K_{jet} = \frac{E}{\eta \ T_{rs}}
\end{equation}

\noindent
where $E$ is the total lobe energy, $\eta$ an efficiency and $T_{rs}$ the
age of the radio source. The total energy of the lobes is derived from the
minimum energy density condition, as discussed in Section 3.1 of
\scite{miley80}, which depends on the redshift of the source,
its dimensions and the flux density at a given frequency. For 3C277.3 we
derive a minimum energy density of $\sim$ 6.1$\times$10$^{-12}$ erg~\pcc,
and so a total energy for the lobes $E \sim$ 6.3$\times$10$^{58}$ erg.

To estimate the age of the radio source, we assume hotspot advance
velocities in the range $0.01c$ --- $0.1c$ (e.g. \pcite{scheuer95}). Thus, for
constant expansion speed, and taking into account that the lobes expand in
two directions, the dynamical age of the radio source will be in
the range 1.6$\times$10$^{4}$ years kpc$^{-1}$ --- 1.6$\times$10$^{5}$
years kpc$^{-1}$. 

The radio axis of 3C277.3 may not be parallel to the plane of the sky, and
the real size of the radio source would then be $D_{real} =
\frac{D_{proj}}{cos \ i}$, where $D_{proj}$ is the radio size projected on
the plane of the sky, and $i$ is the angle between the radio axis and the
plane of the sky.  However, since the uncertainty in the hotspot advance
speed is greater than that in $cos \ i$, we assume for these calculations
$i=0^{\circ}$.  For the adopted cosmology, $D_{proj}$ = 96 kpc (see
Table~\ref{proptab}), thus the age of 3C277.3 is in the range
1.5$\times$10$^{6}$ yr $\lesssim T_{rs} \lesssim$ 1.5$\times$10$^{7}$~yr.

Substituting in formula~(\ref{eq:jet}), assuming $\eta$=0.5
\cite{rawlings91}, we obtain a total kinetic power for the jet in the
range 2.7$\times$10$^{44}$ erg~s$^{-1} \lesssim K_{jet} \lesssim$
2.7$\times$10$^{45}$ erg~s$^{-1}$. Comparing the jet power with the total
emission-line luminosity observed in the region of the radio jet knot,
which is of the order $L_{tot} \sim$ 5.4$\times$10$^{42}$ erg~s$^{-1}$
\footnote{Assuming Case B recombination \cite{osterbrock89}, the total
emission-line luminosity is approximately 20$\times$L(H$\alpha$).}, it can
be seen that efficiencies of the order of $\sim$ 10$^{-2}$ are sufficient
to produce the observed emission-line luminosity.  Thus, it is entirely
possible that jet-induced shocks provide the energy needed to produce the
observed line luminosity in the radio jet knot region.

\medskip

If line emission in the radio jet knot is indeed induced by shocks driven
by interactions between the radio jet and the ambient gas, how can the
quiescent kinematics of this presumably shocked region be explained?  One
possibility is that the shocked gas is not observable as line emission,
and what is detected is the precursor gas which lies around the shocked
structures and which is photoionized by the UV photons emitted by the
shocked gas behind the shock front. The precursor region is also
characterized by a high ionization state \cite{dopita95}, being therefore
consistent with the observations (see Section~\ref{res:ionizcoma}).  But
why is the shocked gas not observable?

\medskip

{\bf(a)} We first consider that the shocked material may not have had enough
time to cool down, therefore line radiation from the shocked gas would
not yet have been emitted.

In order to estimate the cooling time ($t_{cool}$) behind a strong shock
we use the following formula, which has been derived from that given in
Section~2.3 of \scite{klein94}:

\begin{equation}
\label{eq:1}
t_{cool} = 3.3 \times 10^{4} \left(\frac{{v_{c}}^{3}}{n_{0}}\right) \ \rm s,
\end{equation}

\noindent 
where $v_{c}$ is the velocity of the shock through the clouds (in units of
\kms) and $n_{0}$ is the density of the gas ahead of the shock (in units
of \pcc).  Assuming that what is observed is the precursor unshocked gas,
the measured densities of the gas in this region give $n_{0}$ = 170 ---
300 \pcc \ (\pcite{clark96,vanbreugel85}, respectively). Thus, for shock
velocities through the clouds $v_{c}$ = 200 --- 500 \kms, we obtain
cooling times in the range $t_{cool} \approx$ 30 --- 770 years, which is
much less than the time the clouds had to cool down ($\gtrsim$10$^{6}$
years) \footnote{Note that if the region is a multi-phase molecular cloud
complex, the densities could be much lower in some parts of the region
interacting with the jet, in which case the cooling times would be much
longer, and the low density gas may not yet have cooled.}.

Consequently, the shocked gas in the region of the knot has had enough
time to cool down since the initial passage of the shock front, and thus
it should be detectable in emission lines. Therefore, the lack of
disturbed kinematics in the knot cannot be explained in this way, unless
this cloud has only just wandered into the jet, or it is being
continuously shock-heated by the passage of the jet.

\medskip

{\bf(b)} Another possibility is that the shocked clouds are destroyed
before they cool down and emit line radiation. 

Calculations by Klein et al. (1994)\nocite{klein94} show that the
destruction time of a shocked cloud is several cloud crushing times, where
the cloud crushing time is the characteristic time for the shock to cross
through the cloud and is defined by

\begin{equation}
\label{eq:3}
t_{cc} \equiv \chi^{\frac{1}{2}} \ \frac{a_{\circ}}{v_{s}} 
\end{equation}

\noindent
where $\chi$ is the ratio of the density of the cloud $\rho_{c}$ to that
of the intercloud medium $\rho_{h}$, $a_{\circ}$ is the initial cloud
radius, and $v_{s}$ is the velocity of the shock in the intercloud
medium. Assuming that the clouds are in pressure equilibrium with the hot
ambient gas, then $\rho_{c} v_{c}{^2} \approx \rho_{h} v_{s}^{2}$. Thus
the cloud crushing time can be expressed in terms of the shock velocity
through the clouds:

\begin{equation}
\label{eq:4}
t_{cc} \approx \frac{a_{\circ}}{v_{c}}
\end{equation}

\noindent
Then, for the clouds to be destroyed before they cool down, we require
$t_{cc}<t_{cool}$. Combining formulas (\ref{eq:1}) and (\ref{eq:4}), the
following constraint on the initial radius of the cloud is obtained:

\begin{equation}
\label{eq:5}
a_{\circ} < 1.07 \times 10^{-9}  \left(\frac{{v_{c}}^{4}}{n_{0}}\right) \
\rm pc,
\end{equation}

\noindent where $v_{c}$ and $n_{0}$ are in units of \kms \ and \pcc,
respectively. For initial cloud densities of the order $n_{0} \approx$ 170
--- 300 \pcc \  and shock velocities through the clouds in the range
$v_{c}$ = 200 --- 500 \kms, we obtain that the initial radii of the
clouds need to be $a_{\circ} <$ 0.4 pc so as to be destroyed before they
cool down.

\medskip
     
Alternatively, if indeed the shocked gas is being seen, one explanation
would be that the clouds were uniformly accelerated by the jet
interactions. This could explain the constant velocity shifts and the
narrow linewidths of this region. However, it would be very surprising if
this were the case for this jet-cloud interaction when usually the
resulting kinematics from such interactions are highly disturbed and
complex.

Another possibility is that large linewidths are not observed in the
region due to a geometrical effect; if the shocked gas is moving parallel
to the plane of the sky, then it cannot not be detected
spectroscopically. However, we consider this possibility extremely
unlikely because if the ambient gas is being highly disturbed by the
passage of the jet, it is difficult to believe that the resulting motions
are exclusively parallel to the plane of the sky, since entrainment of
the gas in the hot post-shock wind and turbulence resulting from the
interactions with the jet would be expected to be three-dimensional.

We conclude that the emission-line region coincident with the radio jet
knot in 3C277.3 does not show highly disturbed kinematics because the
shocked gas in the region may have not yet cooled down and it is not
observable as line emission, either because the ambient clouds are
destroyed before they cool (if $a_{0} < 0.4$ pc), or more likely because
the shocked gas is continuously heated by the passage of the jet. In this
case the line emission from the knot is dominated by the
shock-photoionized precursor emission from the region surrounding the
expanding shock front.

\subsection{Origin of the gas}

A fundamental question regarding the formation and evolution of radio
galaxies which still remains unclear concerns the origin of the extended
gas in these sources.

As has been previously suggested, the angular momentum of the
emission-line nebula associated with radio galaxies may reflect the origin
of the gas in these systems.  Previous studies of powerful low-redshift
radio galaxies report a tendency for the gas rotation axis to align with
the radio axis (\pcite{simkin79,kotanyi79,heckman85,baum92}).
Interestingly, for the galaxies in the current sample, if their underlying
velocity field is due to rotation, it seems that the rotation axis of the
gas and the radio axis tend to be misaligned by more than $\sim$
60$^{\circ}$ [see Figs.~\ref{comavelwidth} (top), \ref{171ratvelwid} (top
right) and \ref{265velwidth} (left)]. 

In the case of 3C265, it is of particular interest that the nearest
companion galaxy (Fig.~\ref{265o2}), aligns along the direction of the
UV/optical emission elongation of 3C265, which is close to the proposed
plane of rotation of the gas.  It has been suggested that the gas and dust
present along this direction is associated with an interaction between
3C265 and its companion (Sol{\'o}rzano-I{\~n}arrea et
al. 2002)\nocite{carmen02}. It thus seems that the plane of rotation of
the emission-line gas would correspond to a gas stream which reflects a
merger or accretion event, at least for 3C265.

If this interpretation of the velocity fields as gas rotation is correct,
then this would suggest that 3C277.3, 3C171 and 3C265 are fundamentally
different to many other sources, and this may then be related to why they
show such strong jet-cloud interactions. For example, if the radio axis
happens to align close to the merger direction, strong jet-cloud
interactions are likely to arise.

Nevertheless, it must not be forgotten that other mechanisms may dominate the
velocity fields of these sources. Certainly large kinematic changes are
seen in the regions of the radio hotspots, and it may be that the rest of
the velocity field of these sources is also dominated by outflows induced
by the radio source. Clearly, more observations are required to analyse in
detail the velocity fields in jet-cloud interaction sources.

\section{Conclusions}

We present the results obtained from integral-field spectroscopic
observations of three powerful radio galaxies (3C277.3, 3C171 and 3C265),
all of which were previously known to be undergoing jet-cloud interactions
to some extent. We map and analyse the morphology, kinematics and
ionization of the emission-line gas in the haloes of these galaxies.

We find that the emission-line regions which are close to the radio
hotspots show disturbed kinematics and low ionization states, as expected
if jet-cloud interactions are taking place in such locations. By
contrast, the region coincident with the radio jet knot in 3C277.3, which
is believed to be undergoing a strong jet-cloud interaction, presents
quiescent kinematics and high-ionization states. We discuss several
possibilities to explain this result.

We show for the first time that the effects of the radio source can extend
far from the radio jet axis. Our images of 3C171 and 3C265 indicate that
the lateral expansion of the cocoon has a significant effect in the
kinematics and ionization of the emission-line gas.

We detect the presence of a stellar-photoionized HII region in the
extended emission-line nebula of 3C277.3.
  
In addition, if the underlying velocity fields of the gas are due to
rotation, we find that the gas rotation axis and the radio axis tend to be
misaligned. This disagrees with previous studies of nearby radio galaxies,
and suggests that the galaxies of our sample may be fundamentally
different to many other sources.

\section*{Acknowledgments}

This work is based on observations made at the Observatorio del Roque de
los Muchachos, La Palma, Spain. This research has made use of the
NASA/IPAC Extragalactic Database (NED) which is operated by the Jet
Propulsion Laboratory, California Institute of Technology, under contract
with the National Aeronautics and Space Administration.  CSI acknowledges
support from PPARC. We thank Raffaella Morganti and Martin Hardcastle for
providing the radio maps of 3C277.3 and 3C171, respectively.  CSI thanks
Bego{\~n}a Garc{\'\i}a-Lorenzo from the ING for her help with the INTEGRAL
reduction package. CSI particularly thanks Philip Best for a careful
reading of the manuscript, many constructive discussions and
encouragement.  We thank the anonymous referee for useful comments and
suggestions which helped to improve this article.

\bibliographystyle{mnras}
\bibliography{reference}

\end{document}